\documentclass[a4paper,11pt]{article}

\usepackage{jheppub} 
\usepackage{tikz}
\usetikzlibrary{intersections,calc,arrows.meta}

\allowdisplaybreaks

\usepackage{amsmath}
\usepackage{amssymb}
\usepackage{latexsym}
\usepackage{mathtools}
\usepackage{comment}
\usepackage[mathscr]{euscript}

\usepackage{bm}
\usepackage{subcaption}

\hypersetup{colorlinks,citecolor=TokiwaIro,linkcolor=RuriIro,urlcolor=RuriIro,bookmarksopen,bookmarksnumbered,pdfstartview=FitH,linktocpage}
\definecolor{RuriIro}{rgb}{0.,0.28,0.60}
\definecolor{TokiwaIro}{rgb}{0.,0.39,0.16}

\newcommand{\nn}{\nonumber}
\newcommand{\del}{\partial}
\newcommand{\mc}{\mathcal}

\renewcommand{\Re}{\mathop{\mathrm{Re}}}
\renewcommand{\Im}{\mathop{\mathrm{Im}}}

\newcommand{\diag}{\mathop{\mathrm{diag}}}

\newcommand{\sgn}{\mathop{\mathrm{sgn}}}

\definecolor{dgreen}{rgb}{0.2,0.51,0.19}

\newcommand{\MPl}{M_{\mathrm{Pl}}}
\newcommand{\DBI}{\mathrm{DBI}}
\newcommand{\Ei}{\mathop{\mathrm{Ei}}\nolimits}

\makeatletter
\gdef\@fpheader{}
\makeatother

\title{
Black Hole Extremality in Nonlinear Electrodynamics: A Lesson for Weak Gravity and Festina Lente Bounds
}

\author[a]{Yoshihiko Abe,}
\author[b,c]{Toshifumi Noumi,}
\author[b,c]{and Kaho Yoshimura}

\affiliation[a]{Department of Physics, University of Wisconsin-Madison, Madison, WI 53706, USA}
\affiliation[b]{Graduate School of Arts and Sciences, University of Tokyo, Komaba, Meguro-ku, Tokyo 153-8902, Japan}
\affiliation[c]{Department of Physics, Kobe University, Kobe 657-8501, Japan}

\emailAdd{yabe3@wisc.edu}
\emailAdd{tnoumi@g.ecc.u-tokyo.ac.jp}
\emailAdd{yoshimura-kaho848@g.ecc.u-tokyo.ac.jp}

\preprint{KOBE-COSMO-23-06,
UT-Komaba/23-4}

\abstract{
We study black hole extremality in nonlinear electrodynamics motivated by the Weak Gravity Conjecture (WGC) and the Festina Lente (FL) bound. For illustration, we consider the Euler-Heisenberg model and the Dirac-Born-Infeld model in asymptotically flat spacetime, de Sitter spacetime, and anti-de Sitter spacetime. We find that in all cases the extremal condition enjoys a certain monotonicity expected by the WGC. This provides evidence for the conjecture beyond the leading order corrections to the Einstein-Maxwell theory. We also study how light charged particles modify the mass-charge relation of Nariai black holes in de Sitter spacetime and discuss possible implications for the FL bound. Besides, we point out an interesting similarity between our black hole analysis and gravitational positivity bounds on scattering amplitudes.

}

\begin{document} 
\setcounter{tocdepth}{2}
\maketitle
\flushbottom

\section{Introduction}

Thermodynamic properties of black holes play a central role in the study of quantum gravity. In the context of the Swampland Program~\cite{Vafa:2005ui}, more specifically, various thought experiments on charged black holes have been performed to explore possible quantum gravity constraints on the charged particle spectrum. See, e.g.,~\cite{Palti:2019pca,vanBeest:2021lhn,Agmon:2022thq} for review articles.

\medskip
A famous example for such swampland conditions is the Weak Gravity Conjecture (WGC)~\cite{Arkani-Hamed:2006emk}, which predicts existence of {\it a charged state} whose mass-to-charge ratio is smaller than unity in an appropriate unit. The conjecture is equivalent to requiring that all the black holes have to decay unless they are protected by (super)symmetries. Applying this to macroscopic extremal black holes, which have zero temperature and thus cannot decay through the standard Hawking radiation mechanism, implies existence of the WGC state.

\begin{figure}[t]
 \centering
 \includegraphics[width=0.4\textwidth]{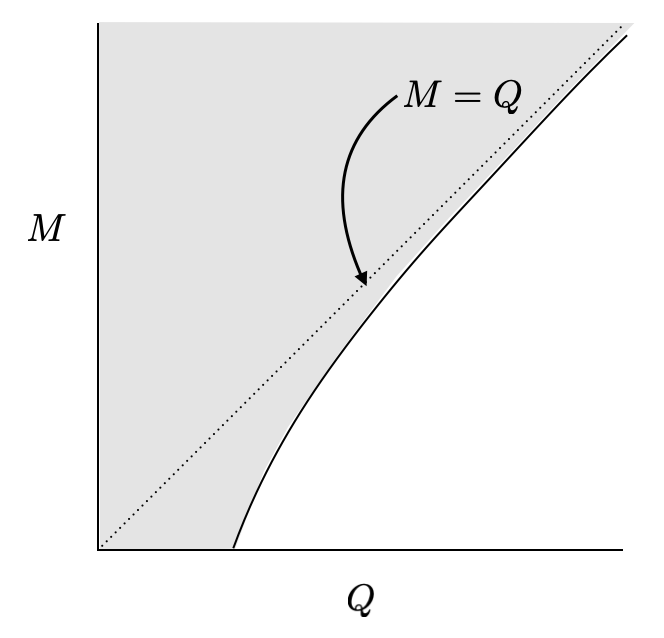}
 \caption{
Typical string theory spectrum: the gray region is populated by charged states and its boundary (the extremal curve) monotonically approaches to the $M=Q$ line, where $M$ and $Q$ are mass and charge in an appropriate unit.}
 \label{fig:WGCtower}
\end{figure}

\medskip
While the original conjecture requires {\it a single WGC state}, various generalizations have been explored and studied in the past decade (see, e.g.,~\cite{Harlow:2022gzl} for a review). In particular, the subLattice/Tower WGC~\cite{Heidenreich:2015nta, Heidenreich:2016aqi, Montero:2016tif, Andriolo:2018lvp} predicts existence of an infinite tower of WGC states at various energy scales. Indeed, known string theory examples accommodate a tower of WGC states both below and above the Planck scale, and the extremal curve satisfies a certain monotonicity as depicted in Fig.~\ref{fig:WGCtower}. This background picture behind the WGC has been confirmed, e.g., by studying string theory compactification~\cite{Arkani-Hamed:2006emk,  Lee:2018urn,Lee:2019tst,Klaewer:2020lfg},  modular invariance~\cite{Heidenreich:2016aqi,Montero:2016tif,Aalsma:2019ryi} and higher derivative corrections to the black hole extremality~\cite{Natsuume:1994hd, Kats:2006xp,Cheung:2018cwt,Hamada:2018dde,Bellazzini:2019xts, Charles:2019qqt, Jones:2019nev, Loges:2019jzs,Goon:2019faz, Cano:2019oma, Cano:2019ycn,Cremonini:2019wdk, Chen:2020rov, Loges:2020trf,  Bobev:2021oku, Arkani-Hamed:2021ajd, Cremonini:2021upd, Aalsma:2021qga,Ma:2021opb,Noumi:2022ybv}.

\begin{figure}[t]
 \centering
 \includegraphics[width=0.5\textwidth]{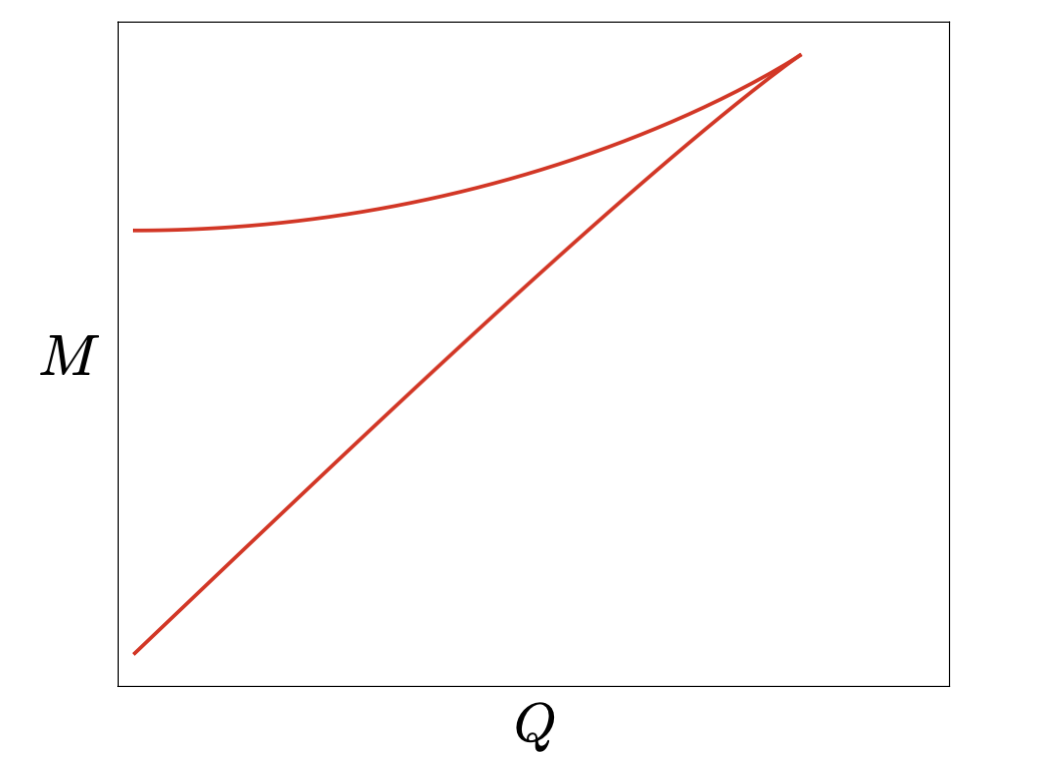}
 \caption{In de Sitter spacetime, charged black holes populate the finite region surrounded by the red curve, where $M$ and $Q$ are mass and charge in an appropriate unit.}
 \label{fig:intro_sharkfin}
\end{figure}

\medskip
More recently, an interesting Swampland condition called the Festina Lente (FL) bound was proposed based on thought experiments about charged black holes in de Sitter spacetime~\cite{Montero:2019ekk}. As depicted in Fig.~\ref{fig:intro_sharkfin}, black holes in de Sitter space have an upper bound on the mass for a given charge to fit inside the cosmological horizon. By postulating that the black holes saturating the bound (Nariai black holes) decay into neutral black holes, rather than naked singularities, a lower bound $m\gtrsim \sqrt{gqM_{\rm Pl}H}$ on the mass of charged particles was proposed~\cite{Montero:2019ekk}, where $m$ and $q$ are the mass and charge of the particle, and $g$, $M_{\rm Pl}$ and $H$ are the gauge coupling, the reduced Planck mass and the Hubble constant. In contrast to the WGC, the FL bound has to be satisfied by all the charged particles to avoid fast discharge processes that lead to naked singularities. See, e.g.,~\cite{Montero:2021otb,Lee:2021cor,Ban:2022jgm} for its  phenomenological implications.

\begin{figure}[t]
 \centering
 \includegraphics[width=\textwidth]{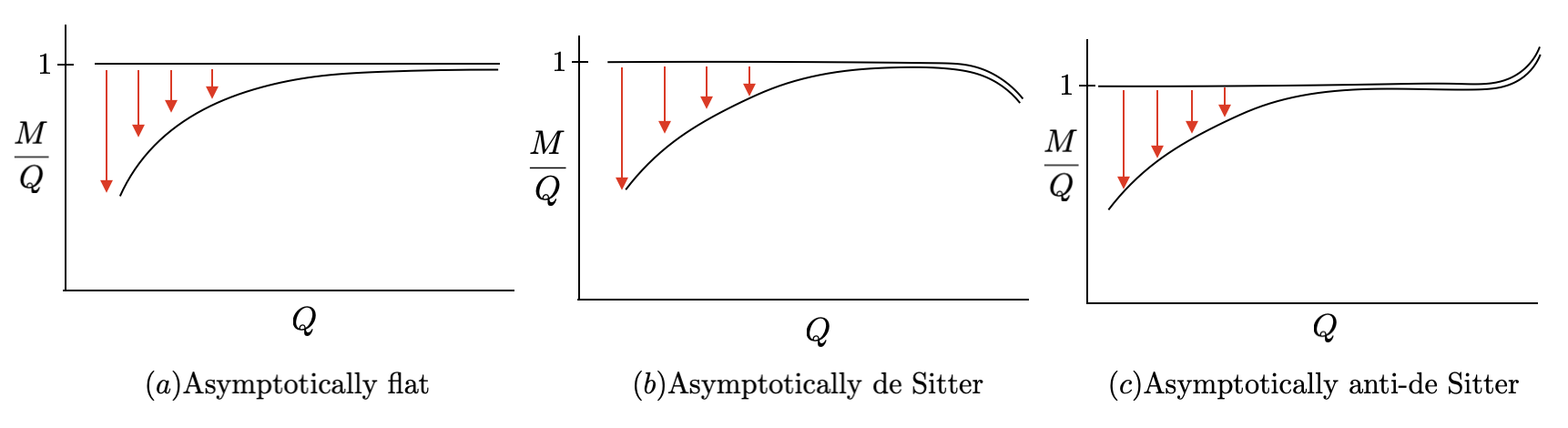}
 \caption{
In flat space, the WGC implies that the mass-to-charge ratio of extremal black holes are lowered by quantum corrections and the correction is monotonic with respect to the charge $Q$. We expect similar monotonicity in dS and AdS as well\protect\footnotemark[1].}
 \label{fig:monotonicity_image}
\end{figure}

\medskip
In this paper, following these developments, we study the black hole extremality in nonlinear electrodynamics in asymptotically
flat spacetime, de Sitter spacetime, and anti-de Sitter spacetime. Our motivation is two-sided: One is in confirming the monotonicity of quantum corrections to the black hole extremality beyond derivative expansions. See Fig.~\ref{fig:monotonicity_image}. In the literature, a lot of evidences for the monotonicity are collected about leading order corrections to the Einstein-Maxwell theory~\cite{Natsuume:1994hd, Kats:2006xp,Cheung:2018cwt,Hamada:2018dde,Bellazzini:2019xts, Charles:2019qqt, Jones:2019nev, Loges:2019jzs,Goon:2019faz, Cano:2019oma, Cano:2019ycn,Cremonini:2019wdk, Chen:2020rov, Loges:2020trf,  Bobev:2021oku, Arkani-Hamed:2021ajd, Cremonini:2021upd, Aalsma:2021qga,Ma:2021opb,Noumi:2022ybv}. While such lower dimensional operators are theoretically well controlled, the derivative expansion is applicable only for large enough black holes. Our purpose in the present paper is to go beyond the derivative expansion, studying the black hole extremality in the Euler-Heisenberg model (EH) and  Dirac-Born-Infeld (DBI) model as illustrative examples for nonlinear electrodynamics. We confirm the expected monotonicity in both models for all signs of the cosmological constant. In this context we also point out an interesting similarity between our black hole analysis and positivity bounds on scattering amplitudes~\cite{Pham:1985cr,Adams:2006sv}, especially in the presence of gravity (see~\cite{Hamada:2018dde,Bellazzini:2019xts,Alberte:2020jsk,Tokuda:2020mlf,Herrero-Valea:2020wxz,Caron-Huot:2021rmr,Alberte:2021dnj,Bellazzini:2021oaj,Caron-Huot:2022ugt,Chiang:2022jep,Herrero-Valea:2022lfd,deRham:2022gfe,Noumi:2022wwf,Hamada:2023cyt} for recent developments).

\footnotetext[1]{See, e.g., Refs.~\cite{Antoniadis:2020xso, Nakayama:2015hga,Benakli:2021fvv} for extension of the WGC to dS and AdS.}
\medskip
The other motivation is in the FL bound. The FL bound was proposed by postulating that Nariai black holes do not decay into naked singularities. While the original argument about discharge processes through the Schwinger mechanism highly depend on the charged black hole spectrum, light charged particles may modify the black hole solutions. Thus, it is of great interests how light charged particles saturating the FL bound modify the black hole solutions and more specifically the Nariai curve. Based on this motivation, we use the Euler-Heisenberg model to study how light charged particles modify the Nariai curve. Interestingly, we find for magnetic black holes that the Nariai curve is flattened in the presence of light charged particles. This motivates further studies of the Schwinger mechanism of Nariai black holes and sharpening the FL bound.

\medskip
This paper is organized as follows:
In Sec.~\ref{sec:2}, we first review the charged black hole solutions of the Einstein-Maxwell theory and introduce a general procedure to calculate the mass-to-charge ratio of extremal and Nariai black holes in nonlinear electrodynamics. 
In Sec.~\ref{sec:3}, we study the extremal curve in the EH model and the DBI model in asymptotically flat spacetime. 
In Sec.~\ref{sec:4} we extend the analysis to the nonzero cosmological constant. In particular, for asymptotically de Sitter case, we study how the Nariai curve is modified by light charged particles and discuss its possible implications for the FL bound. In Sec.~\ref{sec:5} we discuss similarity between our black hole analysis and gravitational positivity bounds on scattering amplitudes.
Then we conclude the paper in Sec.~\ref{sec:6} with an outlook for future work. Some technicalities and our notation are collected in Appendices.

\section{Generality}
\label{sec:2}

In this section we review charged black hole solutions in the Einstein-Maxwell theory and then present general construction of charged black holes in nonlinear electrodynamics.

\subsection{Einstein-Maxwell theory}
\label{sec:2.1}

Consider the Einstein-Maxwell theory in four dimensions with a cosmological constant $\Lambda$:
\begin{align}
\label{eq:EinsteinMaxwellaction}
    S=\int d^4x\sqrt{-g}\left[\frac{1}{16\pi G}(R-2\Lambda)-
     \frac{1}{4g_e^2} F_{\mu\nu} F^{\mu\nu}
     \right]\,,
\end{align}
where $g_e$ is the (electric) gauge coupling constant and the Newton constant $G$ is related to the reduced Planck mass
$\MPl$ as $G = \frac{1}{8 \pi \MPl^2}$. We use $G$ and $\MPl$ interchangeably depending on the context for notational simplicity. 
The Einstein equation is
\begin{align}
\label{eq:Einsteineq}
	G_{\mu\nu}  =8\pi GT_{\mu\nu}- g_{\mu\nu}\Lambda\,,
\end{align}
where $G_{\mu\nu} \coloneqq R_{\mu\nu} - \frac{1}{2}  g_{\mu\nu}R$ is the Einstein tensor and $T_{\mu\nu}$ is the energy-momentum tensor of the matter sector.
In the Einstein-Maxwell theory, the energy-momentum tensor reads 
\begin{align}
	T_{\mu\nu}
    =\frac{1}{g_e^2} \left( F_{\mu\lambda} F_\nu{}^\lambda - \frac{1}{4} g_{\mu\nu} F_{\rho\sigma} F^{\rho\sigma} \right)\,.
\end{align}
The Maxwell equation and the Bianchi identity are 
\begin{align}
	\label{eq:Maxwelleq1}
	\nabla_{\mu}F^{\mu\nu}=0\,,
	\quad
	\nabla_{\mu}\tilde{F}^{\mu\nu}=0\,,
\end{align}
where the dual field strength is defined by $\tilde{F}^{\mu\nu} \coloneqq \frac{1}{2} \epsilon^{\mu\nu\rho\sigma} F_{\rho\sigma}$.
$\epsilon^{\mu\nu\rho\sigma}$ is the Levi-Civita tensor on the curved spacetime normalized such that $\epsilon^{0123}=-(-g)^{-1/2}$. In other words, it is related to the Levi-Civita symbol $\varepsilon^{\mu\nu\rho\sigma}$ ($\varepsilon^{0123}=-1$ ) as $  \epsilon^{\mu\nu\rho\sigma}=(-g)^{-1/2}\varepsilon^{\mu\nu\rho\sigma}$.
For details of the anti-symmetric tensors and symbols, see Appendix~\ref{app:def-of-epsilon}.

\medskip
We consider static and spherically symmetric black holes with either electric or magnetic charges. Let us employ the following ansatz of the metric in the polar coordinates\footnotemark[2]\footnotetext[2]{ We have $T^t{}_t=T^r{}_r$ in the setups studied in the present paper, so that it is compatible with the Einstein equation to assume $g_{tt} g_{rr} = -1$.}:
\begin{align}
\label{eq:sphericallysymsol}
    ds^2=-f(r)dt^2+\frac{dr^2}{f(r)}+r^2\left(d\theta^2+\sin^2\theta d\phi^2\right)\,.
\end{align}
Then, for magnetic black holes, the Gauss law says 
\begin{align}
    F = \frac{1}{2}F_{\mu\nu}dx^{\mu}\wedge dx^{\nu}=\frac{n\sin\theta }{2} d\theta\wedge d\phi~,
\end{align}
where $n$ is the quantized integer charge of the black hole. Without loss of generality, we assume $n\geq0$ throughout the paper. Note that this field configuration is not modified by higher derivative corrections. The magnetic flux on the two-dimensional sphere reads
\begin{align}
    \int_{S^2} F =\int_{S^2}\frac{n\sin\theta }{2} d\theta\wedge d\phi=2 \pi n\,.
\end{align}
On the other hand, for electric black holes, the Maxwell equation says
\begin{align}
\label{F_electric_BH}
    \frac{1}{2}F_{\mu\nu}dx^{\mu}\wedge dx^{\nu}=k_e\frac{n}{r^2} dr\wedge dt
    =\frac{g_e^2}{4\pi}\frac{n}{r^2}dr\wedge dt\,,
\end{align}
where $k_e=g_e^2/4\pi$ is the (electric) Coulomb constant and $n$ is the quantized integer charge. 
The electric flux on the two-dimensional sphere reads
\begin{align}
\label{ele_charge}
	\int_{S^2} \ast_4 F =g_e^2n\,,
\end{align}
where $\ast_4$ is the four-dimensional Hodge star. More explicitly,
\begin{align}
	\ast_4 F = \frac{\sqrt{-g}}{2! 2!} F^{\mu\nu} \varepsilon_{\mu\nu\rho\sigma} dx^\rho \wedge d x^\sigma.
\end{align}
In contrast to the magnetic case, the field configuration~\eqref{F_electric_BH} is modified in the nonlinear electrodynamics, essentially because definition~\eqref{ele_charge} of the electric charge is modified in the presence of higher derivative corrections.

\medskip
It is also instructive to note
\begin{align}
\frac{1}{4g_e^2} F_{\mu\nu} F^{\mu\nu}= - \frac{g_e^2 n^2}{32 \pi^2 r^4} = - \frac{k_e n^2}{8\pi r^4}  
\quad
\text{for electric black holes}
\end{align}
and
\begin{align}
\frac{1}{4g_e^2} F_{\mu\nu} F^{\mu\nu}= \frac{g_m^2 n^2}{32 \pi^2 r^4} =  \frac{k_m n^2}{8\pi r^4}  
\quad
\text{for magnetic black holes},
\end{align}
which manifests the electric-magnetic duality. Here we introduced the magnetic gauge coupling $g_m=\frac{2\pi}{g_e}$ and the magnetic Coulomb constant $k_m=\frac{g_m^2}{4\pi}=\frac{\pi}{g_e^2}$. Especially in figures, we sometimes parameterize the black hole charge in the unit of gauge couplings as
\begin{align}
Q_e:=g_en\,,
\quad
Q_m:=g_m n\,.
\end{align}
In this paper, we discuss electric and magnetic black holes separately, so that we suppress the subscripts $e,m$ of $g_{e,m}$, $k_{e,m}$, and $Q_{e,m}$ to simplify the notation as $g$, $k$, and $Q$, as long as it is obvious from the context.

\medskip
With the above gauge field configurations, the Einstein equation simply reduces to
\begin{align}
(\theta_r+1)f(r)=1-\frac{Gkn^2}{r^2}-\Lambda r^2\,,
\end{align}
where the Euler operator $\theta_r:=r\frac{\partial}{\partial r}$ counts the exponent of $r$. $f(r)$ is then determined as
\begin{align}
	f(r) =  1 - \frac{2 GM}{r} + \frac{Gk n^2}{r^2} - \frac{\Lambda}{3} r^2\,,
\end{align}
where the integration constant $M$ is identified with the black hole mass.

\subsubsection{Black hole extremality}
\label{subsec:EM_extremality}

Location of the black hole horizon and the cosmological horizon is determined by the equation $f(r)=0$. This means that when a horizon is located at $r=r_H$, the black hole mass $M$ is written as a function of $r_H$ and $n$:
\begin{align}
    \label{eq:MaxwellM}
	M(r_H, n) = \frac{r_H}{2 G} + \frac{kn^2}{2 r_H} - \frac{\Lambda}{6 G} r_H^3\,.
\end{align}
When there exist multiple horizons, the extremal condition for horizon degeneracy reads
\begin{align}
\frac{\del M(r_H , n) }{ \del r_H} = 0\,.
\end{align}

\paragraph{Asymptotically flat geometry.}
First, let us consider the asymptotically flat geometry, i.e., $\Lambda =0$.
Generically, there are two positive real solutions for $f(r)=0$ corresponding to the Cauchy horizon and the event horizon. When the two horizons degenerate, its location $r_H$ is determined by
\begin{align}
	\frac{\del M(r_H, n)}{\del r_H} = \frac{1}{2 G} - \frac{k n^2}{2 r_H^2} = 0\,,
\end{align}
so that we have $r_H = \sqrt{G k} n$.
Substituting this back into Eq.~\eqref{eq:MaxwellM} gives the  extremal condition on flat space:
\begin{align}
GM^2=kn^2\,.
\end{align}

\paragraph{Asymptotically de Sitter geometry.}

Next we consider asymptotically de Sitter (dS) geometry, i.e., $\Lambda>0$. Generically, $f(r)=0$ has three positive real solutions corresponding to the two black hole horizons and the cosmological horizon. The condition for the horizon degeneracy reads
\begin{align}
\label{degenerate_dS}
	\frac{\del M(r_H, n)}{\del r_H} = \frac{1}{2 G} - \frac{k n^2}{2 r_H^2} -\frac{\Lambda}{2G}r_H^2= 0\,.
\end{align}
For $n<(4Gk\Lambda)^{-1/2}$, it has two positive real solutions
\begin{align}
r_{H\pm}=\sqrt{\frac{1\pm \sqrt{1-4Gk\Lambda n^2}}{2\Lambda}}\,.
\end{align}
For $r=r_{H+}$, the cosmological horizon and the black hole event horizon degenerate. Such black holes are called Nariai black holes~\cite{1950SRToh..34..160N,1999GReGr..31..963N}. On the other hand, for $r=r_{H-}$, the two black hole horizons degenerate, corresponding to extremal black holes. Substituting $r_H=r_{H\pm}$ back into Eq.~\eqref{eq:MaxwellM} gives the Nariai condition and the extremal condition, respectively.

\begin{figure}[t]
    \centering
    \includegraphics[width=10cm]{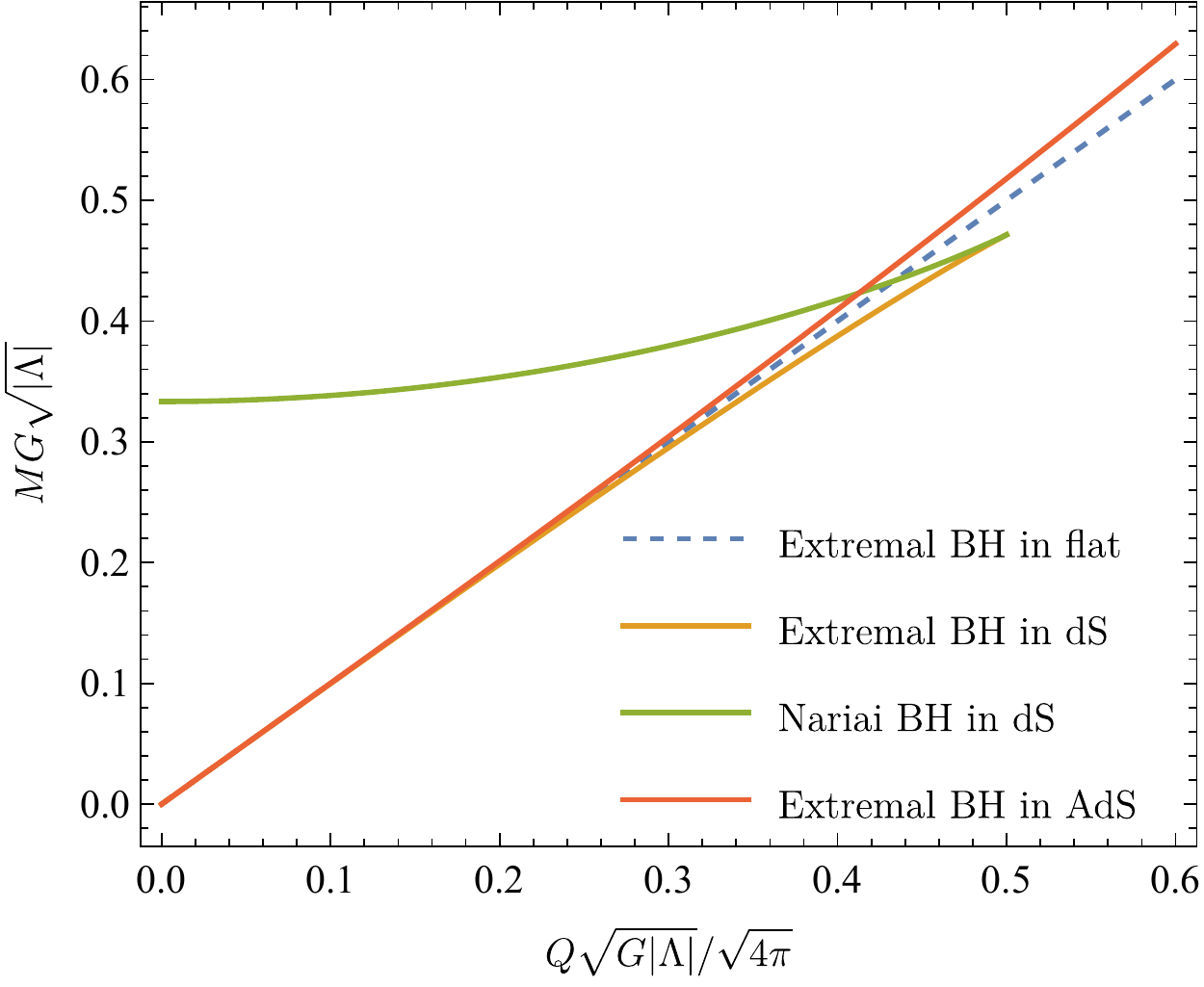}
    \caption{Phase diagram of charged black holes in asymptotically flat, dS, and AdS backgrounds:
    The blue dashed curve is the extremal curve in flat spacetime.
    The orange and green curves are the extremal and Nariai curves in dS, respectively. 
    The red curve is the extremal curve in AdS. We parameterized the charge by $Q=gn$.}
    \label{fig:Maxwell}
\end{figure}

\medskip
When $n=(4Gk\Lambda)^{-1/2}$, the equation~\eqref{degenerate_dS} has only one positive real solution,
\begin{align}
r_H=\sqrt{\frac{1}{2\Lambda}}\,,
\end{align}
for which the three horizons degenerate, corresponding to the ultracold black hole. Note that for $n>(4Gk\Lambda)^{-1/2}$, there is only one positive real solution for $f(r)=0$, corresponding to the cosmological horizon. Therefore, in dS, there exists an upper bound $n\leq(4Gk\Lambda)^{-1/2}$ on the black hole charge. See Fig.~\ref{fig:Maxwell}.

\paragraph{Asymptotically anti-de Sitter geometry.}

Finally, we consider asymptotically anti-de Sitter (AdS) geometry, i.e., $\Lambda<0$. Generically, $f(r)=0$ has two positive real solutions corresponding to the two black hole horizons. The condition~\eqref{degenerate_dS} for the horizon degeneracy has only one positive real solution,
\begin{align}
r_H=\sqrt{\frac{1+ \sqrt{1+4Gk|\Lambda| n^2}}{2\Lambda}}\,,
\end{align}
corresponding to extremal black holes. Substituting this back into Eq.~\eqref{eq:MaxwellM} gives the extremal condition in AdS. See also Fig.~\ref{fig:Maxwell}.

\subsection{Nonlinear electrodynamics}

Let us move on to black holes in nonlinear electrodynamics with the following action\footnotemark[3]\footnotetext[3]{See, e.g.,~\cite{Pellicer:1969cf,Demianski:1986wx,Breton:2003tk,Breton:2007bza,Kruglov:2017mpj,Ayon-Beato:2000mjt,Ayon-Beato:1998hmi,Ayon-Beato:1999qin,Ayon-Beato:1999kuh,Bronnikov:2000vy,Ayon-Beato:2004ywd,Fernando:2016ksb,Fan:2016hvf,Chinaglia:2017uqd,Bronnikov:2017sgg,Rodrigues:2018bdc,Ali:2018boy,Poshteh:2020sgp,Villani:2021lmo,Mehdipour:2021ipf,Kruglov:2021mfy,Lousto:1988sp,Campanelli:1994sj,Nomura:2020tpc,Nomura:2021efi} for earlier works on charged black holes in nonlinear electrodynamics. In particular, our presentation on the construction follows Ref.~\cite{Nomura:2021efi}.}:
\begin{align}
\label{eq:action}
    S=\int d^4x\sqrt{-g}\left[\frac{1}{16\pi G}(R-2\Lambda)+\mathcal{L}(\mathcal{F},\mathcal{G})\right],
\end{align}
where $\mathcal{L}(\mathcal{F},\mathcal{G})$ is a function of $\mathcal{F}$ and $\mathcal{G}$ defined by
\begin{align}
\label{eq:mathcalFG}
	\mc{F} \coloneqq \frac{1}{4} F_{\mu\nu} F^{\mu\nu}\,,
    \qquad
    \mc{G} \coloneqq \frac{1}{4} F_{\mu\nu} \tilde{F}^{\mu\nu} = \frac{1}{8} \epsilon^{\mu\nu\rho\sigma} F_{\mu\nu} F_{\rho \sigma}\,.
\end{align}
To separate the cosmological constant from the gauge field sector, we assume $\mathcal{L}(0,0)=0$. We also assume that $\mathcal{L}(\mathcal{F},\mathcal{G})$ is analytic at $\mathcal{F}=\mathcal{G}=0$ and the Lagrangian is parity invariant.
The Einstein equation reads
\begin{align}
G_{\mu\nu}=8\pi GT_{\mu\nu}- g_{\mu\nu}\Lambda
\end{align}
with the energy-momentum tensor,
\begin{align}
\label{EM-tensor}
	T_{\mu\nu} = - \frac{\del \mc{L}(\mc{F},\mc{G})}{\del \mc{F}} F_{\mu\lambda} F_{\nu}{}^{\lambda} + g_{\mu\nu} \left[ \mc{L(\mc{F},\mc{G})} - \frac{\del \mc{L}(\mc{F},\mc{G})}{\del \mc{G}} \mc{G} \right].
\end{align}
The equation of motion for the Maxwell field and the Bianchi identity are
\begin{align}
\label{eq:Maxwell}
	\nabla_\mu \left[ \frac{\del \mc{L}(\mc{F},\mc{G})}{\del \mc{F}} F^{\mu\nu} + \frac{\del \mc{L}(\mc{F},\mc{G})}{\del \mc{G}} \tilde{F}^{\mu\nu} \right] = 0\,,
	\quad 
	\nabla_\mu \tilde{F}^{\mu\nu} = 0\,.
\end{align}
Note that $\mc{L} = - \frac{1}{g_e^2}\mc{F}$ in the Einstein-Maxwell theory. Eq.~\eqref{eq:Maxwell} provides a nonlinear extension of the Maxwell equation. Below we present general construction of static and spherically symmetric charged black hole solutions in the nonlinear electrodynamics.

\subsubsection{Magnetic black holes}

We begin by magnetic black holes. Again we employ the following ansatz of the metric:
\begin{align}
\label{eq:sphericallysymsol2}
    ds^2=-f(r)dt^2+\frac{dr^2}{f(r)}+r^2\left(d\theta^2+\sin^2\theta d\phi^2\right)\,.
\end{align}
As we mentioned in the previous subsection, the gauge field configuration is unchanged from the Einstein-Maxwell case due to the Gauss law and the charge quantization:
\begin{align}
\label{eq:magneticBH}
    \frac{1}{2}F_{\mu\nu}dx^\mu\wedge dx^\nu=\frac{n \sin\theta}{2} d\theta \wedge d\phi\,,
\end{align}
where $n$ is the quantized integer charge.
We then have
\begin{align}
\label{eq:magneticFG}
    \mathcal{F}=\frac{n^2}{8r^4}, \quad \mathcal{G}=0\,.
\end{align}
Then, the Einstein equation reduces to
\begin{align}
(\theta_r+1)f(r)=1+8\pi G r^2\mathcal{L}\left(\tfrac{n^2}{8r^4},0\right)-\Lambda r^2\,,
\end{align}
which determines $f(r)$ as
\begin{align}
\label{eq:fmagnetic}
	f(r) = 1 - \frac{2 G M}{r} - \frac{\Lambda}{3}r^2 + \frac{8 \pi G}{r} \int_\infty^r dr' r'^2 \mc{L}\left(\tfrac{n^2}{8r'^4},0\right)\,.
\end{align}
Here the integration constant $M$ is again interpreted as the black hole mass. Note that in the large $r$ regime $r\to\infty$, the last term is subdominant compared to the first three terms because we assumed that $\mathcal{L}(\mathcal{F},\mathcal{G})$ is analytic and vanishes at $\mathcal{F}=\mathcal{G}=0$.

\medskip
To identify the extremal condition, it is convenient to express the black hole mass as a function of the horizon radius $r_H$  and the charge $n$:
\begin{align}
\label{eq:Mmagnetic}
    M(r_{H}, n ) = \frac{r_H}{2G} - \frac{\Lambda}{6 G}r_H^3 + 4 \pi \int_\infty^{r_H} dr r^2 \mc{L}\left(\tfrac{n^2}{8r^4},0\right)\,.
\end{align}
Then, the condition for horizon degeneracy reads
\begin{align}
\label{eq:extremalcond_magnetic}
    \frac{\partial M(r_H,n)}{\partial r_H}=\frac{1}{2 G}-\frac{ \Lambda}{2 G}r_H^2+4\pi r_H^2 \mathcal{L}\left(\tfrac{n^2}{8r_H^4},0\right)=0\,.
\end{align}

\subsubsection{Electric black holes}
\label{subsec_nonlinearE}

\paragraph{Legendre transformation.}

Next we consider electric black holes. In contrast to the magnetic case, higher derivative operators modify definition of the electric charge and the Gauss law accordingly. To handle this modification systematically, it is convenient to perform a Legendre transformation of the form,
\begin{align}
\label{eq:Legendre}
\mathcal{H}&=\frac{1}{2 g_e^2}P^{\mu\nu}F_{\mu\nu}-\mathcal{L}
\,,
\quad
P_{\mu\nu}=2 g_e^2 \frac{\partial \mc{L}}{\partial F^{\mu\nu}}\,.
\end{align}
An explicit form of the two-form field $P_{\mu\nu}$ conjugate to $F_{\mu\nu}$ is
\begin{align}
\label{eq:P}
	P_{\mu\nu}=g_e^2 \left[\frac{\del \mc{L}(\mc{F},\mc{G})}{\del \mc{F}} F_{\mu\nu} + \frac{\del \mc{L}(\mc{F},\mc{G})}{\del \mc{G}} \tilde{F}_{\mu\nu}\right]\,,
\end{align}
so that the equation of motion~\eqref{eq:Maxwell} corresponding to the modified Gauss law is simply
\begin{align}
\nabla_{\mu}P^{\mu\nu}=0\,.
\end{align}
We find that in terms of $P_{\mu\nu}$ the equation of motion takes the same form as the standard Gauss law and also it does not depend on the choice of the function $\mathcal{L}(\mathcal{F},\mathcal{Q})$ explicitly. This is why the Legendre transformation~\eqref{eq:Legendre} makes the analysis more tractable. Also, in terms of $\mathcal{L}(\mc{F},\mc{G})$, the Hamiltonian type operator $\mathcal{H}$ is given by
\begin{align}
\label{eq:Hamiltonian}
    \mathcal{H}
    &=2 \frac{\partial \mathcal{L}(\mc{F},\mc{G})}{\partial \mathcal{F}}\mathcal{F}+2 \frac{\partial \mathcal{L}(\mc{F},\mc{G})}{\partial {\mathcal{G}}}\mathcal{G}-\mathcal{L}(\mc{F},\mc{G})\,.
\end{align}

\paragraph{Inverse Legendre transformation.}

In our analysis, we need to perform the inverse Legendre transformation afterwards. By analogy with $\mc{F}$ and $\mc{G}$, let us introduce
\begin{align}
\label{eq:mathcalPQ}
	\mc{P} \coloneqq \frac{1}{4} P_{\mu\nu} P^{\mu\nu}\,,
	\quad 
	\mc{Q} \coloneqq \frac{1}{4} P_{\mu\nu} \tilde{P}^{\mu\nu} = \frac{1}{8} \epsilon^{\mu\nu\rho\sigma} P_{\mu\nu} P_{\rho\sigma}\,.
\end{align}
If we think of $\mathcal{H}$ as a function of $\mathcal{P}$ and $\mathcal{Q}$, $F_{\mu\nu}$ is given by
\begin{align}
	F_{\mu\nu} (P) &= 2 g_e^2 \frac{\partial \mc{H}}{\partial P^{\mu\nu}}
 \label{eq:F-PPtilde}
 =g_e^2\left[\frac{\partial \mc{H}(\mc{P},\mc{Q})}{\partial \mc{P}}P_{\mu\nu}+\frac{\partial \mc{H}(\mc{P},\mc{Q})}{\partial \mc{Q}}\tilde{P}_{\mu\nu}\right]\,,
\end{align}
and correspondingly the Lagrangian $\mathcal{L}$ reads 
\begin{align}
 \mc{L}&= \frac{1}{2 g_e^2} P^{\mu\nu} F_{\mu\nu} - \mc{H}= 2 \frac{\del \mc{H}(\mc{P},\mc{Q})}{\del \mc{P}} \mc{P} + 2 \frac{\del \mc{H}(\mc{P},\mc{Q})}{\del \mc{Q}} \mc{Q} - \mc{H}(\mc{P},\mc{Q})\,.
\end{align}
We can also write the energy-momentum tensor~\eqref{EM-tensor} in terms of $\mc{H}$, $\mc{P}$, and $\mc{Q}$ as
\begin{align}
	T_{\mu\nu} &=-\frac{\del \mc{L}(\mc{F},\mc{G})}{\del \mc{F}} F_{\mu\lambda}F_{\nu}{}^\lambda + g_{\mu\nu} \left[ \mc{L}(\mc{F},\mc{G}) -  \frac{\del \mc{L}(\mc{F},\mc{G})}{\del \mc{G}} \mc{G} \right]
	\nn \\
	&= - \frac{\del \mc{H}(\mc{P},\mc{Q})}{\del \mc{P}} P_{\mu\lambda} P_{\nu}{}^{\lambda} + g_{\mu\nu} \left[ 2 \frac{\del \mc{H}(\mc{P},\mc{Q})}{\del \mc{P}} \mc{P} + \frac{\del \mc{H}(\mc{P},\mc{Q})}{\del \mc{Q}} \mc{Q} - \mc{H}(\mc{P},\mc{Q}) \right]\,,
	\label{eq:EMtensor-NL-P}
\end{align}
where we used the following identities in four dimensions:
\begin{align}
	F^{\mu\lambda} \tilde{F}_{\nu\lambda} = \delta^\mu_\nu \mc{G}\,,
	\quad 
	P^{\mu\lambda} \tilde{P}_{\nu\lambda} = \delta^\mu_\nu \mc{Q}\,.
\end{align}

\paragraph{Black hole solutions.}

Now we are ready to construct black hole solutions. We employ the static and spherically symmetric ansatz~\eqref{eq:sphericallysymsol2} of the metric and solve the modified Maxwell equations~\eqref{eq:Maxwell} that are written in terms of $P_{\mu\nu}$ and $\mc{H}$ as follows:
\begin{align}
\label{eq:electricMaxwell}
	\nabla_\mu P^{\mu\nu} = 0\,,
   \qquad 
   \nabla_{\mu}\left[\frac{\partial \mc{H}(\mc{P},\mc{Q})}{\partial \mc{P}}\tilde{P}^{\mu\nu}-\frac{\partial \mc{H}(\mc{P},\mc{Q})}{\partial \mc{Q}}P^{\mu\nu}\right]=0\,.
\end{align}
For electric black holes, $P_{\mu\nu}$ is specified by Eq.~\eqref{eq:electricMaxwell} as
\begin{align}
    \frac{1}{2}P_{\mu\nu}dx^\mu\wedge dx^\nu=k_e\frac{n}{r^2}dr\wedge dt
    =\frac{g_e^2}{4\pi}\frac{n}{r^2}dr\wedge dt
\end{align}
with a quantized integer charge $n$. Correspondingly, we have
\begin{align}
    \mathcal{P}=-\frac{g_e^4 n^2}{32\pi^2 r^4}\,, \quad \mathcal{Q}=0\,.
\end{align}
Nonzero components of the energy-momentum tensor \eqref{eq:EMtensor-NL-P} are 
\begin{align}
	T^t{}_t = T^r{}_r &=  - \mc{H}\left(-\tfrac{g_e^4 n^2}{32\pi^2 r^4},0\right)\,,
	\\
	T^\theta{}_\theta = T^\phi{}_\phi &=\left[  2\frac{\del \mc{H}(\mc{P},\mc{Q})}{\del \mc{P}}\mc{P} - \mc{H}(\mc{P},\mc{Q})\right]_{\mc{P}=-\tfrac{g_e^4 n^2}{32\pi^2 r^4},\mc{Q}=0}\,.
\end{align}
Then, the Einstein equation reduces to
\begin{align}
(\theta_r+1)f(r)=1-8\pi G r^2\mathcal{H}\left(-\tfrac{g_e^4 n^2}{32\pi^2 r^4},0\right)-\Lambda r^2\,,
\end{align}
which determines $f(r)$ as
\begin{align}
\label{eq:felectric}
	f(r) = 1 - \frac{2 GM}{r} - \frac{\Lambda}{3} r^2 - \frac{8 \pi G}{r} \int_\infty^r dr' r'^2 \mc{H}\left(-\tfrac{g_e^4 n^2}{32\pi^2 r'^4}, 0\right)\,.
\end{align}
Here the integration constant $M$ is interpreted as the black hole mass. As before, we write the black hole mass  as a function of the horizon radius $r_H$,
\begin{align}
\label{eq:Melectric}
    M(r_{H},n)=\frac{r_{H}}{2G}-\frac{1}{2G}\frac{\Lambda}{3}r_{H}^3-4\pi\int^{r_{H}}_\infty dr r^2 \mathcal{H}\left(-\tfrac{g_e^4 n^2}{32\pi^2 r^4}, 0\right)\,,
\end{align}
which gives the following condition for horizon degeneracy:
\begin{align}
\label{eq:extremalcond_electric}
	\frac{\del M(r_H,  n)}{\del r_H} = \frac{1}{2G} - \frac{r_H^2 \Lambda}{2G} - 4 \pi r_H^2 \mc{H}\left(-\tfrac{g_e^4 n^2}{32\pi^2 r_H^4}, 0\right) = 0\,.
\end{align}
Note that the above results for electric black holes are reproduced by a simple replacement $\mathcal{L}(-\tfrac{n^2}{8 r_H^4}, 0)\to-\mathcal{H}(-\tfrac{g_e^4 n^2}{32\pi^2 r_H^4}, 0)$ in the corresponding magnetic results. In particular the charge relation is $Q_m=g_mn\to Q_e=g_en$.

\section{Asymptotically flat black holes in nonlinear electrodynamics}
\label{sec:3}
In this section we study the extremal condition of asymptotically flat black holes in the nonlinear electrodynamics. For illustration, we consider the Euler-Heisenberg model and the Dirac-Born-Infeld model, and confirm the monotonicity expected by the WGC.

\subsection{Euler-Heisenberg black holes}
\label{sec:3.1}

The Euler-Heisenberg (EH) model~\cite{Heisenberg:1936nmg,Weisskopf:1936hya,Schwinger:1951nm} is an effective field theory (EFT) after integrating out a minimally coupled charged particle at the one-loop level. It is applicable when the electromagnetic fields are nearly constant at the Compton scale of the charged particle. See the last paragraph of this subsection for validity of this approximation.
Also except in Sec.~\ref{sec:5} we assume that gravitational corrections are subdominant.
A concrete form of the effective Lagrangian after integrating out a charged scalar/fermion is\footnotemark[4]\footnotetext[4]{We assume that the charged particle has a unit charge.}
\begin{align}
    \mathcal{L}(\mc{F},\mc{G})=
    \left\{\begin{array}{lcc}
    \displaystyle
    -\frac{\mathcal{F}}{g_e^2}+\frac{1}{32\pi^2}\int^\infty_0 \frac{ds}{s} e^{-sm^2} \left[\frac{\mathcal{G}}{\Im \cosh(sX)}-\frac{1}{s^2}+\frac{\mathcal{F}}{3}\right] &\,\,& (\text{scalar}) \,,  \\[5mm]
    \displaystyle
    -\frac{\mathcal{F}}{g_e^2}-\frac{1}{32\pi^2}\int^\infty_0 \frac{ds}{s} e^{-sm^2}\left[4\frac{\Re \cosh(sX)}{\Im \cosh(sX)}\mathcal{G}-\frac{4}{s^2}-\frac{8}{3}\mathcal{F}\right]
    &\,\,&
    (\text{fermion})\,,
    \end{array}\right.
    \label{eq:EH-Lagrangian-scalar}
\end{align}
where $g_e$ is the electric gauge coupling and $m$ is the mass of the electrically charged particle integrated out. We also introduced
\begin{align}
	X \coloneqq \sqrt{ 2 ( \mc{F} + i \mc{G})}\,.
\end{align}
To ignore the Schwinger effect~\cite{Schwinger:1951nm} and work with static black hole solutions, our EH analysis focuses on magnetic black holes, leaving electric black holes for future work.

\medskip
As we discussed in the previous section, we have $\mc{G}=0$ for magnetic black holes, so that what we practically need is a concrete form of the function $\mathcal{L}(\mc{F},0)$. Noticing
\begin{align}
\label{eq:coshexpand}
	\cosh ( s X) = \cosh (\sqrt{2\mc{F}}  s) + i \frac{ s \sinh(\sqrt{2 \mc{F}} s)}{\sqrt{2 \mc{F}}} \mc{G} + \mc{O}(\mc{G}^2)\,,
\end{align}
we have
\begin{align}
    \mathcal{L}(\mc{F},0)=
    \left\{\begin{array}{lcc}
    \displaystyle
    -\frac{\mathcal{F}}{g_e^2}+\frac{\mc{F}}{32\pi^2}\int^\infty_0 \frac{ds}{s} e^{\frac{-sm^2}{\sqrt{2 \mc{F}}}} \left(\frac{2}{s\sinh s}-\frac{2}{s^2}+\frac{1}{3}\right) &\,\,& (\text{scalar}) \,,  \\[5mm]
    \displaystyle
    -\frac{\mathcal{F}}{g_e^2}-\frac{\mc{F}}{4\pi^2}\int^\infty_0 \frac{ds}{s} e^{\frac{-sm^2}{\sqrt{2 \mc{F}}}} \left(\frac{1}{s\tanh s}-\frac{1}{s^2} - \frac{1}{3}\right)
    &\,\,&
    (\text{fermion})\,,
    \end{array}\right.
    \label{eq:EH-Lagrangian}
\end{align}
where we rescaled the integration variable as $s\to \frac{s}{\sqrt{2\mc{F}}}$. Note that its explicit form up to four-derivatives is given by
\begin{align}
    \mathcal{L}(\mc{F},0)=
    \left\{\begin{array}{lcc}
    \displaystyle
    -\frac{\mathcal{F}}{g_e^2}+\frac{7}{2880\pi^2m^4}\mc{F}^2
    +\mathcal{O}(\mc{F}^3)&\,\,& (\text{scalar}) \,,  \\[5mm]
    \displaystyle
     -\frac{\mathcal{F}}{g_e^2}+\frac{1}{90\pi^2m^4}\mc{F}^2
    +\mathcal{O}(\mc{F}^3)
    &\,\,&
    (\text{fermion})\,,
    \end{array}\right.
    \label{eq:EH-Lagrangian-four-derivative}
\end{align}
which can be used when comparing our full order analysis of the EH black hole with earlier works on the leading order corrections.

\begin{figure}[t]
 \centering
 \includegraphics[width=0.6\textwidth]{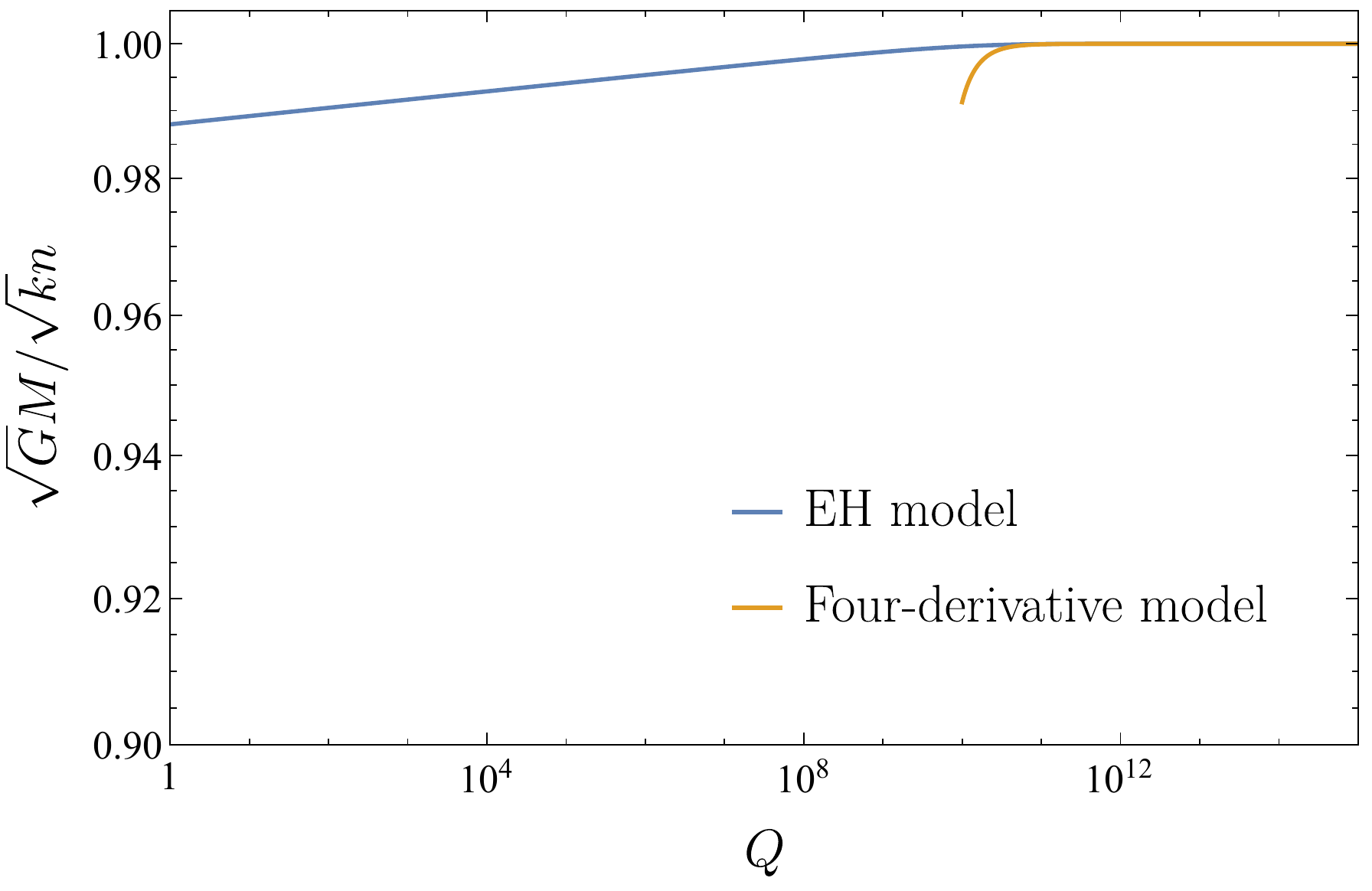}
 \quad
 \includegraphics[width=0.6\textwidth]{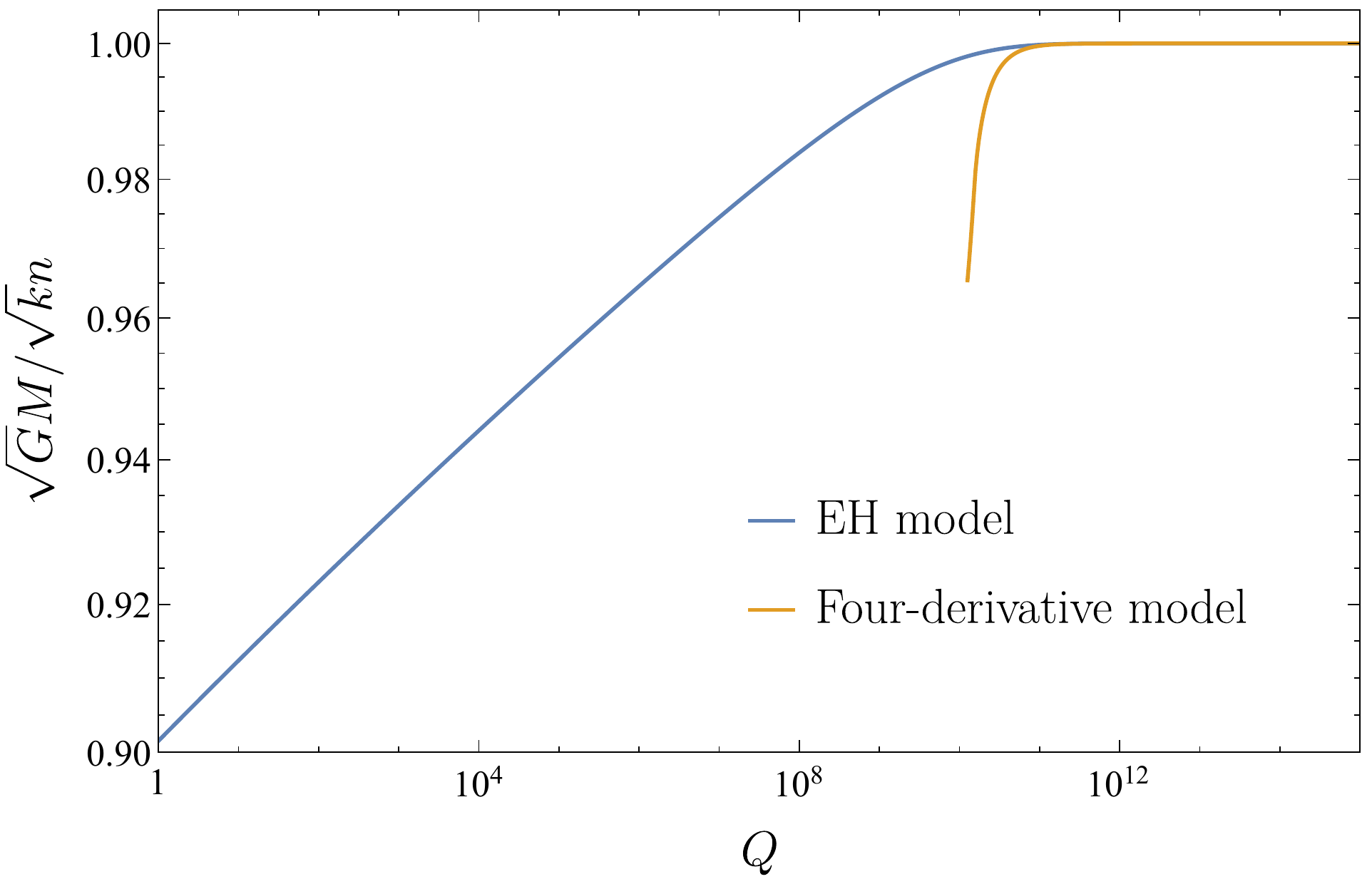}
 \caption{Extremal conditions in the EH model in flat spacetime for $m=10^{-5}\MPl$ and $g_e=1$. The upper/lower figure is for scalar/fermion loop.}
 \label{fig:EH_MtoQ}
\end{figure}

\paragraph{Black hole extremality.}

Now we are ready to determine the black hole extremality. The algorithm to derive the extremal condition is the same as the Einstein-Maxwell case explained in Sec.~\ref{subsec:EM_extremality}: First, we substitute Eq.~\eqref{eq:EH-Lagrangian} and $\Lambda=0$ into the condition~\eqref{eq:extremalcond_magnetic} for horizon degeneracy and solve it for a given magnetic charge $n$ to identify the horizon radius $r_H$ of the extremal black hole. Then, we substitute the obtained $r_H$ into the mass formula~\eqref{eq:Mmagnetic} with Eq.~\eqref{eq:EH-Lagrangian} and $\Lambda=0$, which gives the mass-charge relation of the extremal black hole of the charge $n$. In practice this analysis involves the integral in the EH Lagrangian~\eqref{eq:EH-Lagrangian}, which is difficult to perform analytically. We handle it by using an analytic approximation~\eqref{eq:scEHcalculation}--\eqref{eq:feEHcalculation}. See Appendix~\ref{app:AppendixA} for details. We then numerically solve Eq.~\eqref{eq:extremalcond_magnetic} to obtain $r_H$ and use it to evaluate the mass~\eqref{eq:Mmagnetic}.

\medskip
The mass-to-charge ratio of extremal black holes for $m = 10^{-5} \MPl$ and $g_e=1$ are given in Fig.~\ref{fig:EH_MtoQ}, where the upper/lower panel shows the result for the scalar/fermion case. The blue and orange curves are for the full EH analysis and the four-derivative model (i.e., Eq.~\eqref{eq:EH-Lagrangian-four-derivative} with truncation of the $\mathcal{O}(\mc{F}^3)$ terms), respectively. We confirm the expected monotonicity for both cases. More interestingly, we find that the correction to the extremal condition in the EH model is milder than the four-derivative model, which offers a concept of ultraviolet (UV) completion in the context of black hole extremality. Also note that the orange curve damps rapidly around $Q=g_mn \sim 10^{10}$, beyond which the four-derivative truncation does not work as we discuss at the end of the subsection. It is also useful to notice that the Cauchy horizon disappears at $Q \sim 10^{10}$ for extremal black holes in the model after four-derivative truncation. Therefore, there is no solution for the horizon degeneracy condition~\eqref{eq:extremalcond_magnetic} for $Q \lesssim 10^{10}$. In Sec.~\ref{sec:3.2} we will find a similar phenomenon in the Dirac-Born-Infeld model and give more detailed comments on this point.

\begin{figure}[t]
 \centering
 \includegraphics[width=0.6\textwidth]{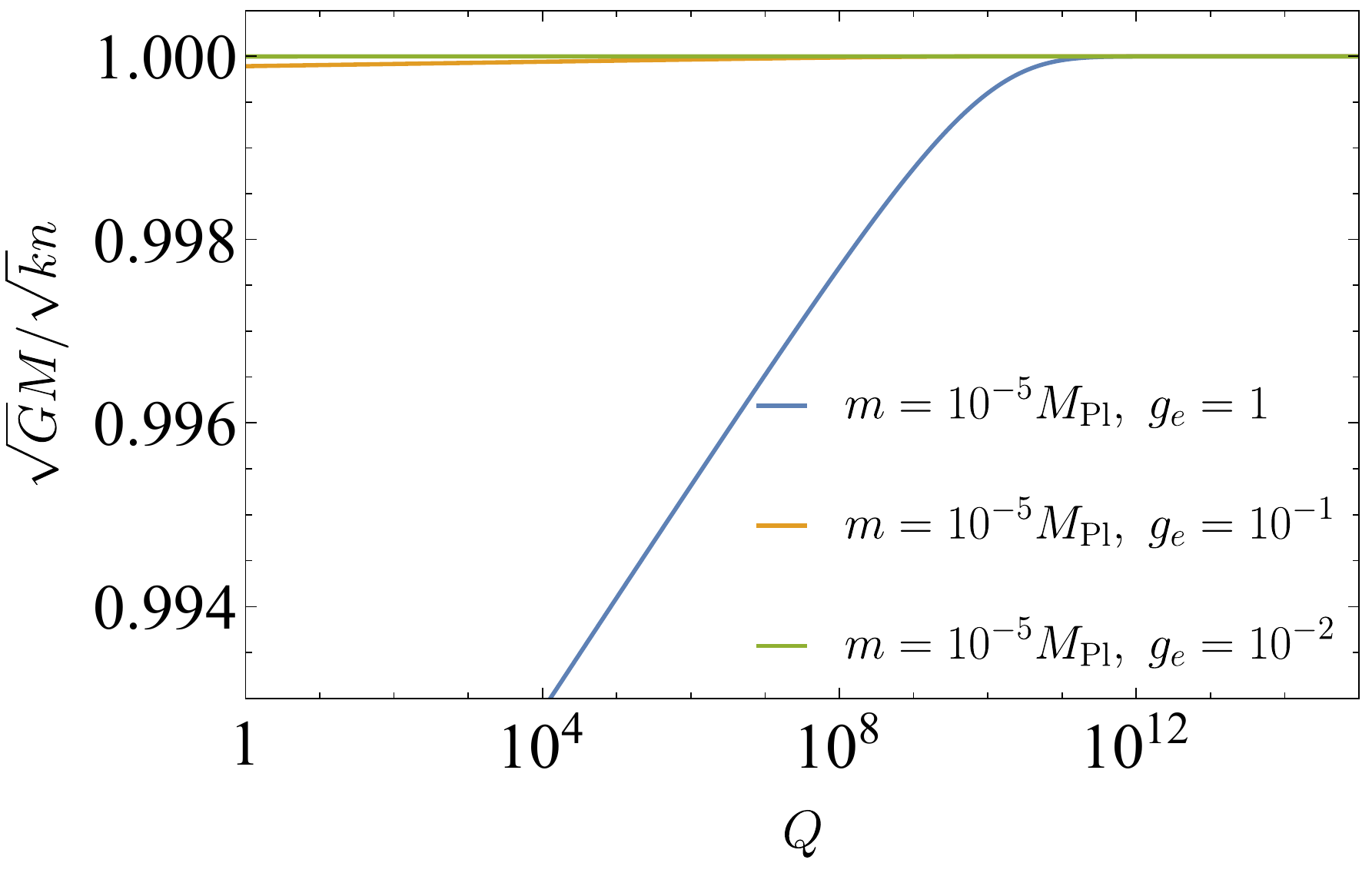}
 \includegraphics[width=0.6\textwidth]{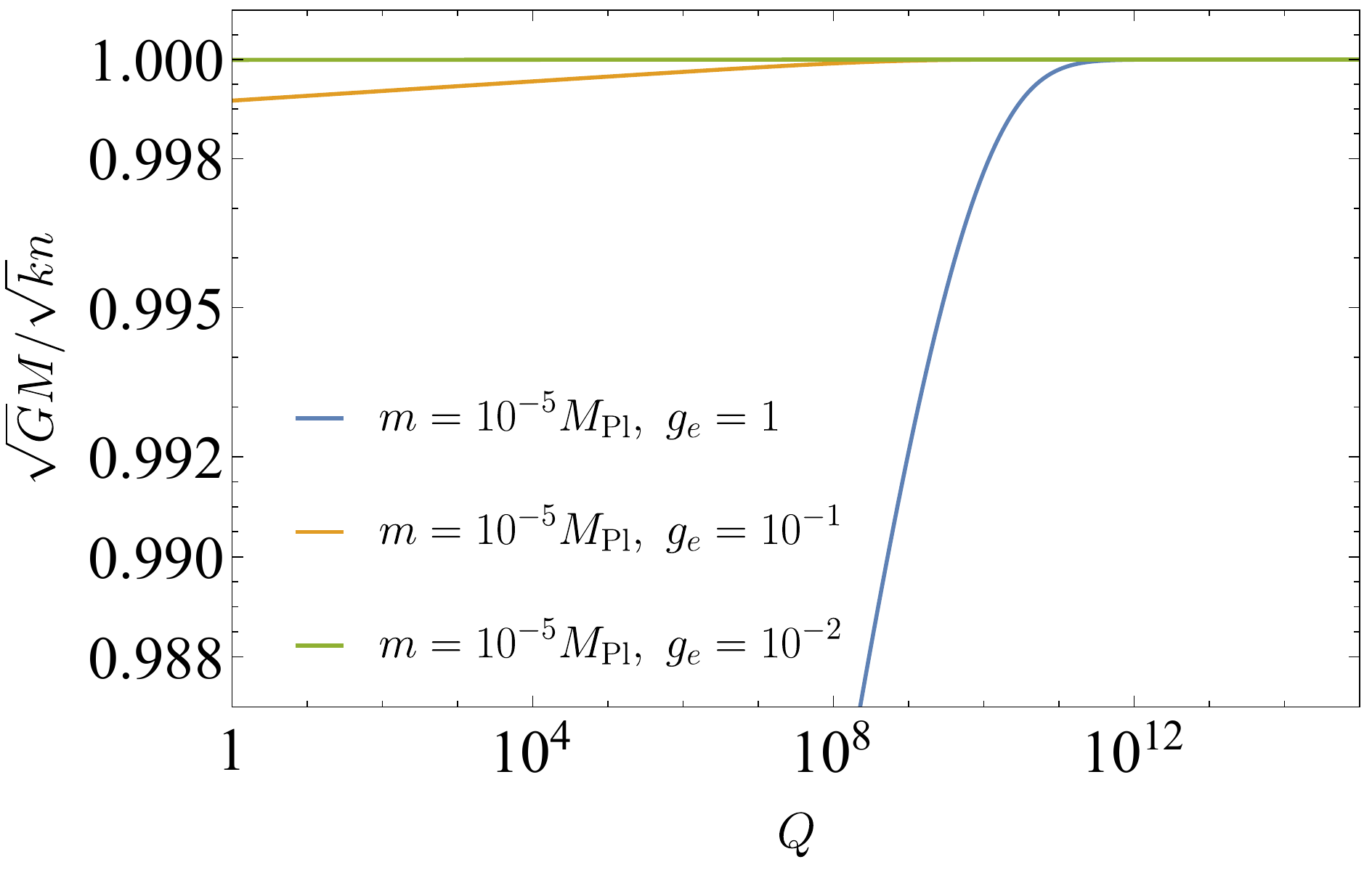}
 \caption{
Gauge coupling dependence of extremal conditions in the EH model in flat spacetime for $m=10^{-5}\MPl$: The upper/lower figure is for scalar/fermion loop. The blue, orange, and green curves are for $g_e=1$, $g_e=10^{-1}$ and $g_e=10^{-2}$, respectively.
 }
 \label{fig:EH_MtoQ_gm}
\end{figure}

\begin{figure}[t]
 \centering
 \includegraphics[width=0.6\textwidth]{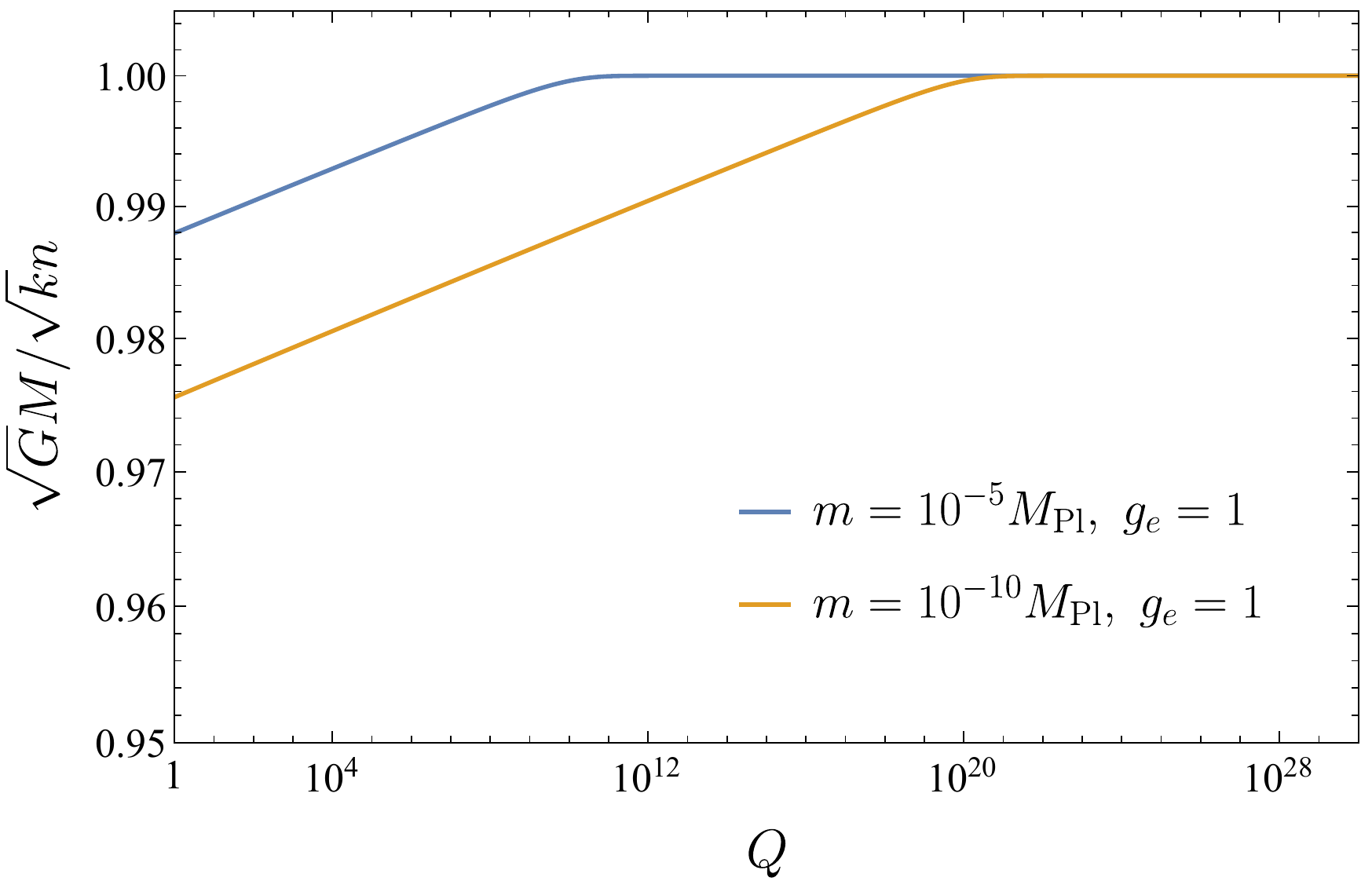}
 \quad
 \includegraphics[width=0.6\textwidth]{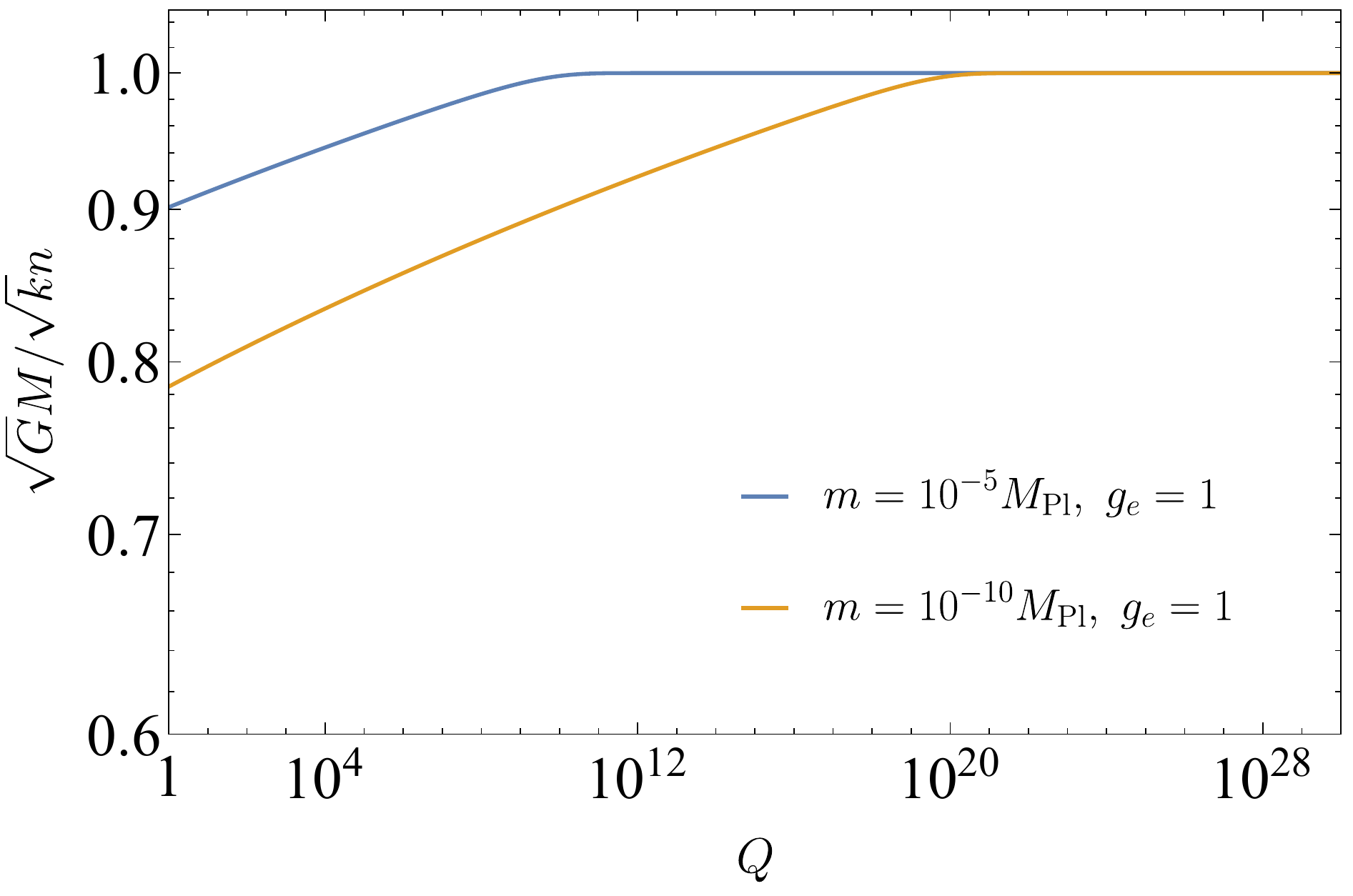}
 \caption{
 Mass dependence of extremal conditions in the EH model in flat spacetime for $g_e=1$: The upper/lower figure is for scalar/fermion loop. The blue and orange curves are for $m=10^{-5}\MPl$ and $m=10^{-10}\MPl$, respectively.
}
 \label{fig:EH_MtoQ_m}
\end{figure}

\medskip
It is also useful to see $g_e$- and $m$-dependence of the extremal condition. Fig.~\ref{fig:EH_MtoQ_gm} shows the $g_e$-dependence of the mass-to-charge ratio of extremal black holes for $m=10^{-5}\MPl$. The upper/lower panel is for the scalar/fermion loop. The correction is larger for a larger electric gauge coupling. Fig.~\ref{fig:EH_MtoQ_m} shows the $m$-dependence for $g_e=1$. Again, the upper/lower panel is for the scalar/fermion loop. The correction is larger for a smaller mass, but the tilt in the small $Q$ region is insensitive to the mass, since the logarithmic behavior is associated with running of the gauge coupling and small $Q$ corresponds to high energy.

\begin{figure}[h]
	\centering
	\begin{tikzpicture}
		\draw[->,thick](-7,0)--(7,0);
		\draw[thick](-3,-0.5)--(-3,0.5);
		\draw[thick](2.8,-0.5)--(2.8,0.5);
		\draw(7.5,0)node{\large$Q$};
		\draw(-5.5,0.8)node{$\frac{1}{m}\left| \frac{\del F}{\del r} \right|\gtrsim |F|$};
		\draw[->,thick](-4,0.8)--(-3.1,0.8);
		\draw[<-,thick](-2.9,0.8)--(-2,0.8);
		\draw(0,0.8)node{Full EH model};
		\draw[->,thick](1.8,0.8)--(2.7,0.8);
		\draw[<-,thick](2.9,0.8)--(3.6,0.8);
		\draw(5,0.8)node{$F^4$ model};
		\draw(-3,-1)node{$\mathcal{O}(\MPl/m)$};
        \draw(2.8,-1)node{$\mathcal{O}\bigl(g_e(\MPl/m)^2 \bigr)$};
	\end{tikzpicture}
	\caption{
The EH model can be used when $Q\gg \MPl/m$ and its full-order analysis beyond the four-derivative approximation is needed for $Q\lesssim g_e(\MPl/m)^2$.
	}
    \label{fig:validEH}
\end{figure}
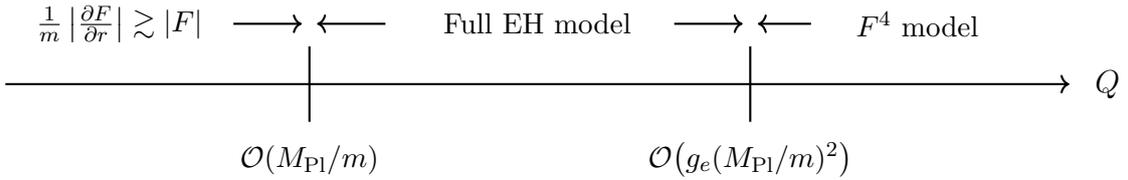

\paragraph{Validity of the Euler-Heisenberg EFT.}

To close the EH analysis, we elaborate on for which charge range the use of the EH Lagrangian~\eqref{eq:EH-Lagrangian-scalar} is justified and the full order analysis without four-derivative truncation is needed. First, when deriving the EH Lagrangian, the electromagnetic field $F_{\mu\nu}$ is assumed to be nearly constant at the Compton scale of the charged particle integrated out, which is schematically given by $\frac{1}{m}\left|\frac{\del F}{\del r}\right|\ll |F|$. For extremal magnetic black holes, this condition is satisfied as long as the horizon radius $r_H\sim g_mn/\MPl\sim Q/\MPl$ is larger than the Compton length $\sim 1/m$ of the charged particle, i.e., $Q\gg \MPl/m$. Also, in the EH model, higher derivative corrections appear schematically in the form, $\mathcal{F}(\mathcal{F}/m^4)^n$ ($n=1,2,\ldots$). Therefore, the derivative expansion does not work and the full-order analysis of the EH model is required when $|\mc{F}|\gtrsim m^4$. For extremal magnetic black holes, this condition reads $n^2/r_H^4\gtrsim m^{-4}$, which is equivalent to $Q\lesssim g_e(\MPl/m)^2$. See Fig.~\ref{fig:validEH} for a summary of the paragraph.

\begin{figure}[t]
 \centering
 \includegraphics[width=0.6\textwidth]{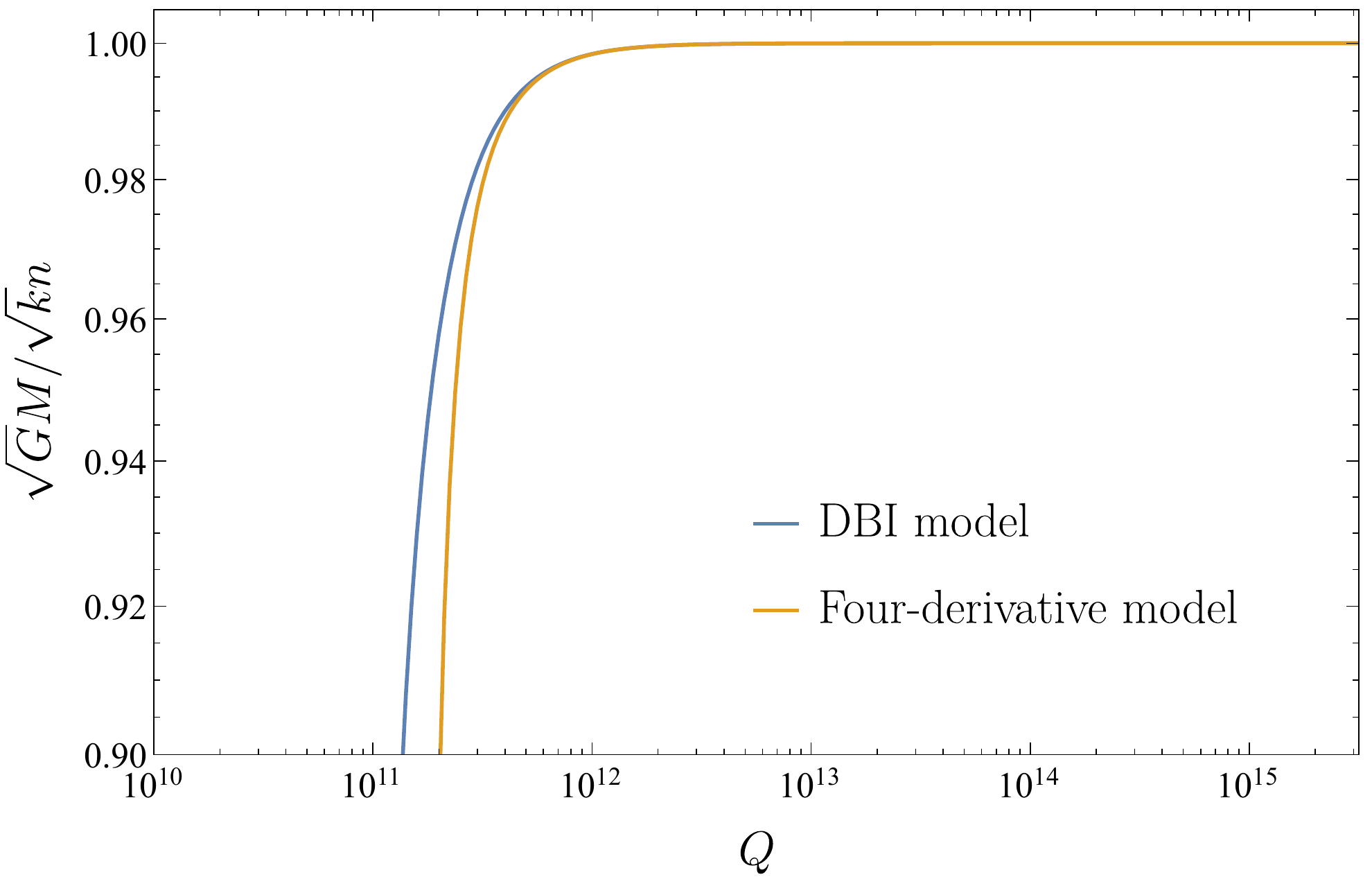}
 \caption{Extremal condition of the DBI model in flat spacetime ($\Lambda_{\rm{DBI}}=10^{-5} \MPl$)
 }
 \label{fig:MtoQDBI}
\end{figure}

\subsection{DBI black holes}
\label{sec:3.2}
The Dirac-Born-Infeld (DBI) model was first introduced to remove divergence associated with the self-energy of charged particles~\cite{Born:1934gh,Dirac:1962iy}. It also provides a low-energy EFT of D-branes, which provides an illustrative example for the nonlinear electrodynamics with a well-motivated UV origin.
In four dimensions, a concrete form of the Lagrangian is
\begin{align}
    \mathcal{L}(\mc{F},\mc{G})=\Lambda_{\rm{DBI}}^4 \left(1-\sqrt{1+\frac{2 \mathcal{F}}{g_e^2\Lambda_{\rm{DBI}}^4}-\frac{ \mathcal{G}^2}{g_e^4\Lambda_{\rm{DBI}}^8}}\,\right)\,,
\end{align}
where $\Lambda_{\rm{DBI}}$ is the brane tension that characterizes the nonlinearity. Note that its derivative expansion up to four-derivatives is
\begin{align}
	\mc{L}(\mc{F},\mathcal{G}) =
 -\frac{\mc{F}}{g_e^2}
 +\frac{1}{2g_e^4\Lambda_\DBI^4}\left(\mc{F}^2+\mc{G}^2\right)+\cdots
	\,,
\end{align}
which we use when comparing our results with the four-derivative analysis in the literature.

\paragraph{Electromagnetic duality.}

As we explained in the previous section, we need a concrete form of $\mathcal{L}(\mc{F},0)$ and $\mathcal{H}(\mc{P},0)$ for the analysis of magnetic and electric black holes, respectively. Performing the Legendre transformation explained in Sec.~\ref{subsec_nonlinearE}, we find
\begin{align}
\label{eq:DBILagrangian}
	\mc{L}(\mc{F},0) = \Lambda_\DBI^4 \left(
		1 - \sqrt{ 1 + \frac{2 \mc{F}}{g_e^2\Lambda_\DBI^4}}\,
	\right)\,,
 \quad
 \mc{H}(\mc{P},0) = -\Lambda_\DBI^4 \left(
		1 - \sqrt{ 1 - \frac{2 \mc{P}}{g_e^2\Lambda_\DBI^4}}\,
	\right)\,.
\end{align}
In particular, $\mc{H}(x,0)=-\mc{L}(-x,0)$ reflects the electromagnetic duality of the DBI model. As we mentioned at the end of Sec.~\ref{subsec_nonlinearE}, this shows that the DBI analysis  for electric black holes is essentially the same as the magnetic black holes, so that we focus on the magnetic case in the following.

\begin{figure}[t]
\begin{tabular}{cc}
\begin{minipage}[b]{0.5\hsize}
\centering
\includegraphics[scale=0.55]{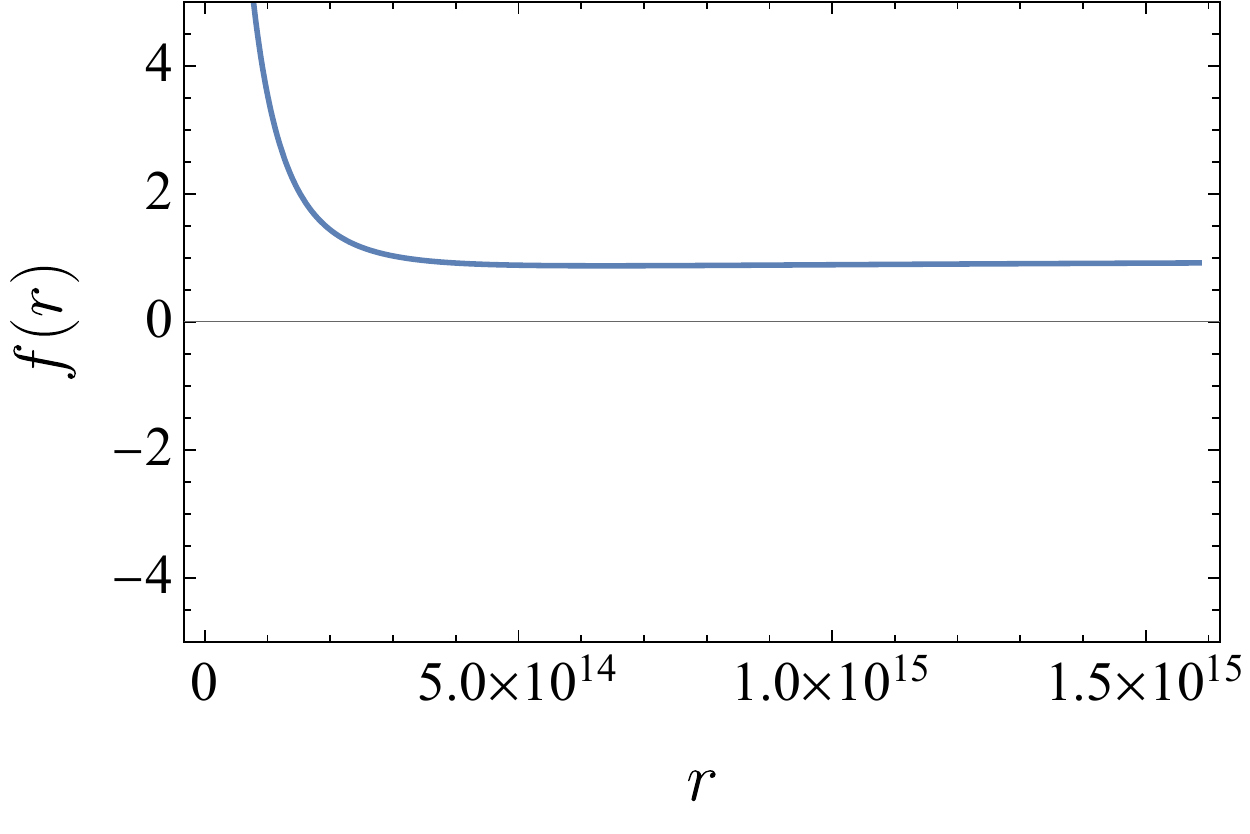}
\subcaption{$Q=4.0\times10^{15},~M=2.0\times10^{15}\MPl$}
\end{minipage}&
\begin{minipage}[b]{0.5\hsize}
\centering
\includegraphics[scale=0.55]{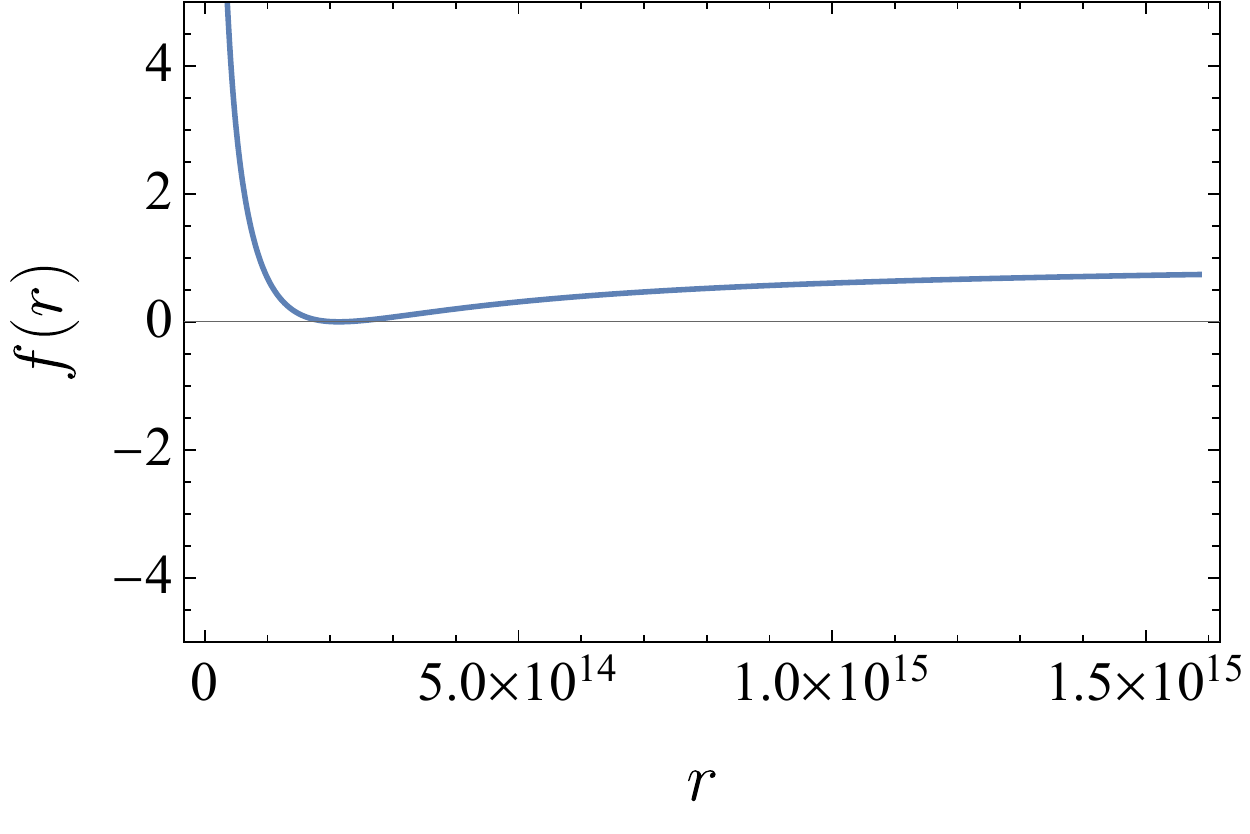}
\subcaption{$Q=4.0\times10^{15},~M\simeq5.6\times10^{15}\MPl$}
\end{minipage}
\\
\\
\begin{minipage}[b]{0.5\hsize}
\centering
\includegraphics[scale=0.55]{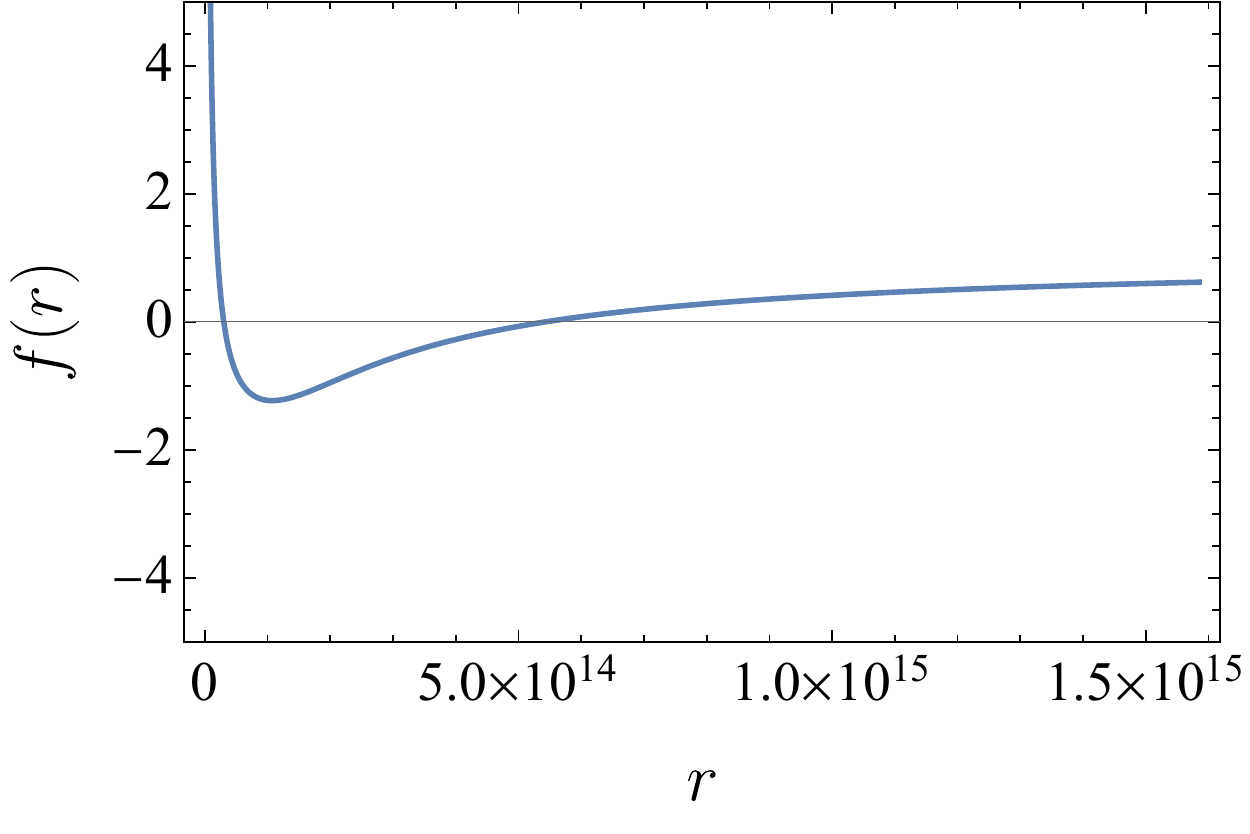}
\subcaption{$Q=4.0\times10^{15},~M=8.0\times10^{15}\MPl$}
\end{minipage} &
\begin{minipage}[b]{0.5\hsize}
\centering
\includegraphics[scale=0.55]{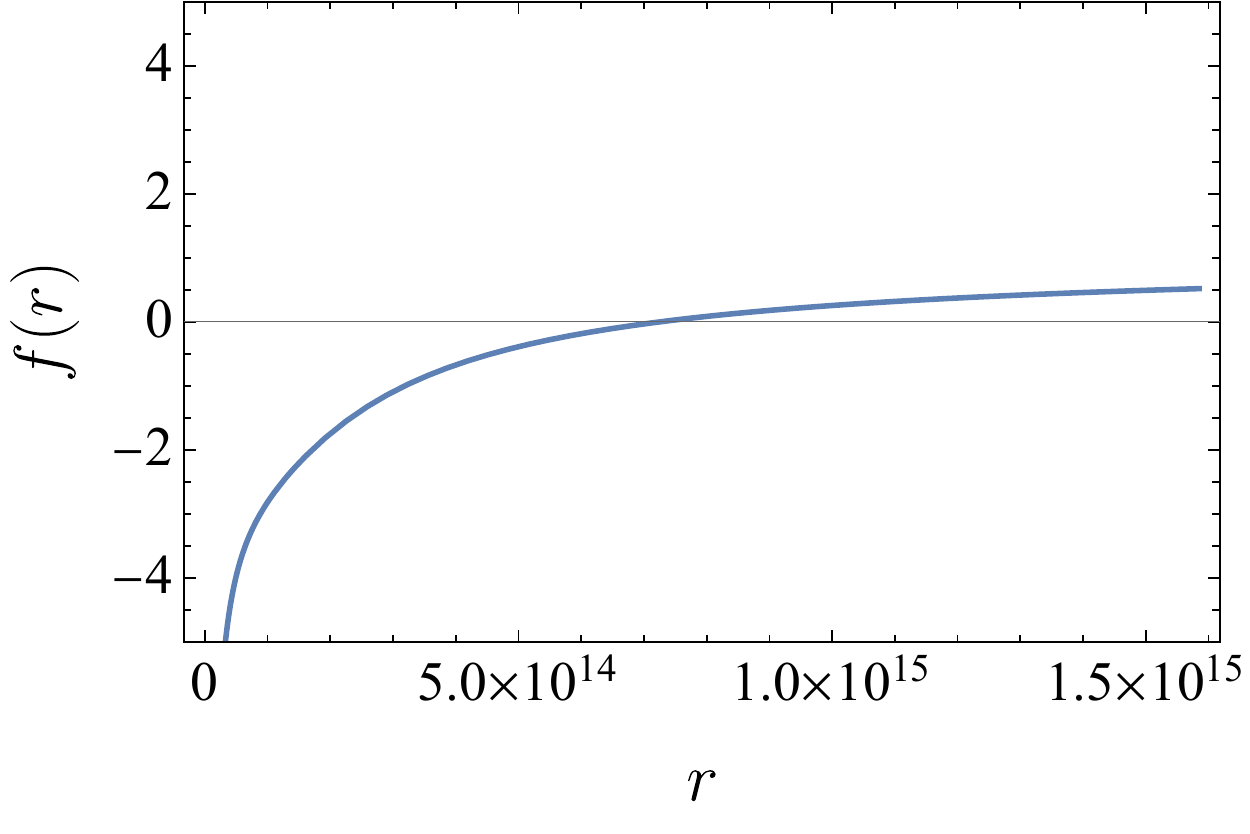}
\subcaption{$Q=4.0\times10^{15},~M=1.0\times10^{16}\MPl$}
\end{minipage}
\\
\\
\begin{minipage}[b]{0.5\hsize}
\centering
\includegraphics[scale=0.55]{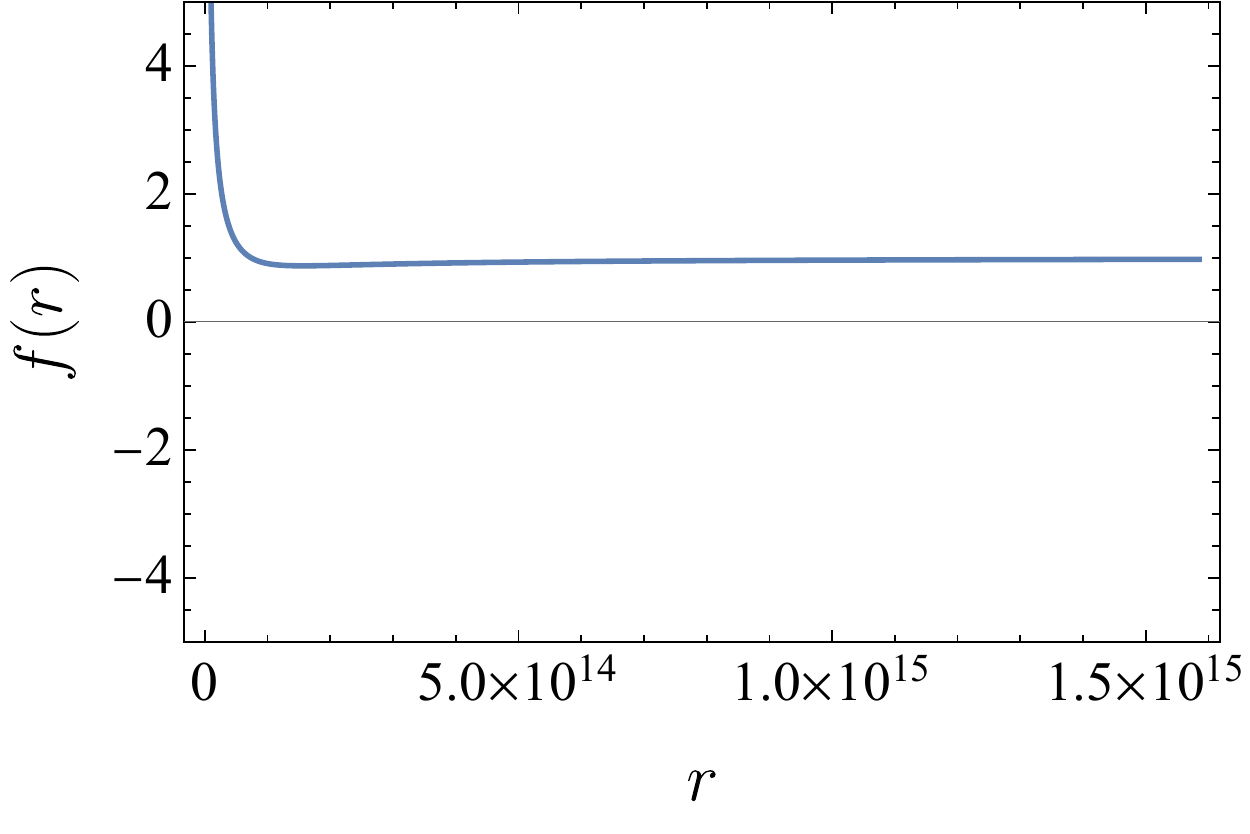}
\subcaption{$Q=1.0\times10^{15},~M=5.0\times10^{14}\MPl$}
\end{minipage}&
\begin{minipage}[b]{0.5\hsize}
\centering
\includegraphics[scale=0.55]{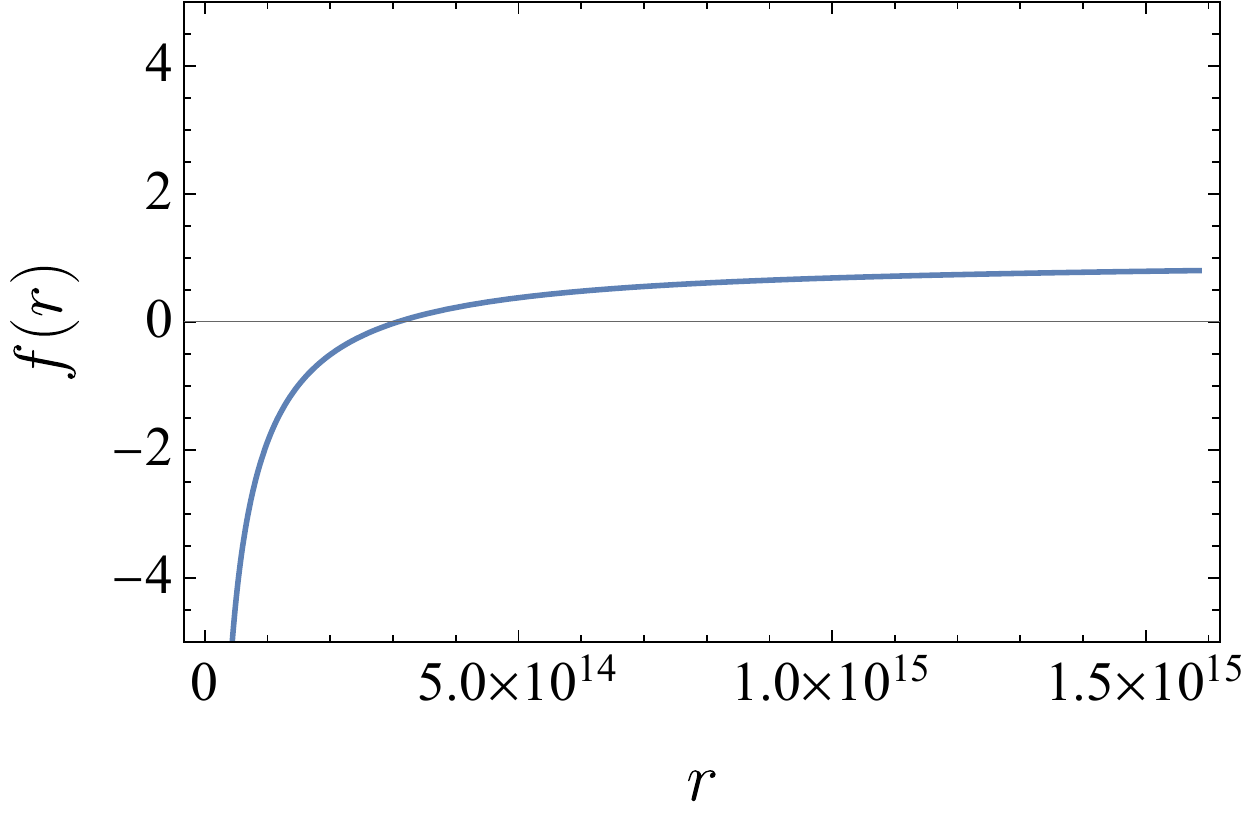}
\subcaption{$Q=1.0\times10^{15},~M=4.0\times10^{15}\MPl$}
\end{minipage}
\end{tabular}
\caption{Typical shapes of $f(r)$ below and above the critical mass and the extremal mass for $\Lambda=10^{-7}\MPl$. For the charge $Q=4.0\times10^{15}$, the critical mass is $M_{\rm crit.}\simeq8.8\times 10^{15}\MPl$ and the extremal mass is $M\simeq5.6\times 10^{15}\MPl$. For the charge $Q=1.0\times 10^{15}$, the critical mass is $M_{\rm crit.}\simeq1.1\times 10^{15}\MPl$ and there is no
degenerate horizon.}
\label{fig: critical_flat}
\end{figure}

\paragraph{Black hole extremality.}

The algorithm to identify the extremal condition is the same as previous examples. First, we solve the condition~\eqref{eq:extremalcond_magnetic} for horizon degeneracy. In the DBI model, the condition reads
\begin{align}
\label{extremal_DBI_general}
\frac{1}{2 G}-\frac{ \Lambda}{2 G}r_H^2+4\pi r_H^2 
\Lambda_\DBI^4 \left(
		1 - \sqrt{ 1 + \frac{Q^2}{16\pi^2\Lambda_\DBI^4r_H^4}}\,
	\right)=0\,,
\end{align}
where we used $Q=g_mn=2\pi n/g_e$ to parameterize the magnetic charge. Also, we kept the cosmological constant $\Lambda$ general for later reference. For the asymptotically flat case $\Lambda=0$, the solution for Eq.~\eqref{extremal_DBI_general} is given by
\begin{align}
\label{rH_DBI_flat}
r_H=
\sqrt{\frac{G}{4\pi }}\sqrt{Q^2-\frac{1}{4G^2 \Lambda_\DBI^4}}\,.
\end{align}
This shows that horizon degeneracy occurs only for $Q\geq (2G\Lambda_\DBI^2)^{-1}$. More comments on this point will be given shortly in the last paragraph of the subsection.

\medskip
Next we evaluate the mass of the extremal black hole for a given charge $n$ using the mass formula~\eqref{eq:Mmagnetic}. In the DBI model, the mass formula reads
\begin{align}
\label{eq:magneticDBIM}
    M=\frac{r_H}{2 G}-\frac{\Lambda}{6 G}r_H^3-\frac{4}{3}\pi  \Lambda_{\DBI}^4 {r_H}^3 \left[\, _2F_1\left(-\frac{3}{4},-\frac{1}{2};\frac{1}{4};-\frac{Q^2}{16 \pi ^2 {r_H}^4 {\Lambda _\DBI}^4}\right)-1\right]\,,
\end{align}
where $_2F_1$ is the Gauss hypergeometric function. Again we kept the cosmological constant $\Lambda$ general for later reference. Substituting Eq.~\eqref{rH_DBI_flat} and $\Lambda=0$ into Eq.~\eqref{eq:magneticDBIM} gives the extremal condition. Fig.~\ref{fig:MtoQDBI} shows the extremal curve of the DBI model with $\Lambda_{\rm{DBI}}=10^{-5} \MPl$ (blue curve), where we confirm the expected monotonicity. Also, we find that the correction to the extremal condition in the DBI model is milder than the four-derivative model (orange curve). Similarly to the EH case, this shows that the DBI model provides a UV completion of the four-derivative model.

\paragraph{Phase structure of horizons.}

\begin{figure}[t]
 \centering
 \includegraphics[width=0.7\textwidth]{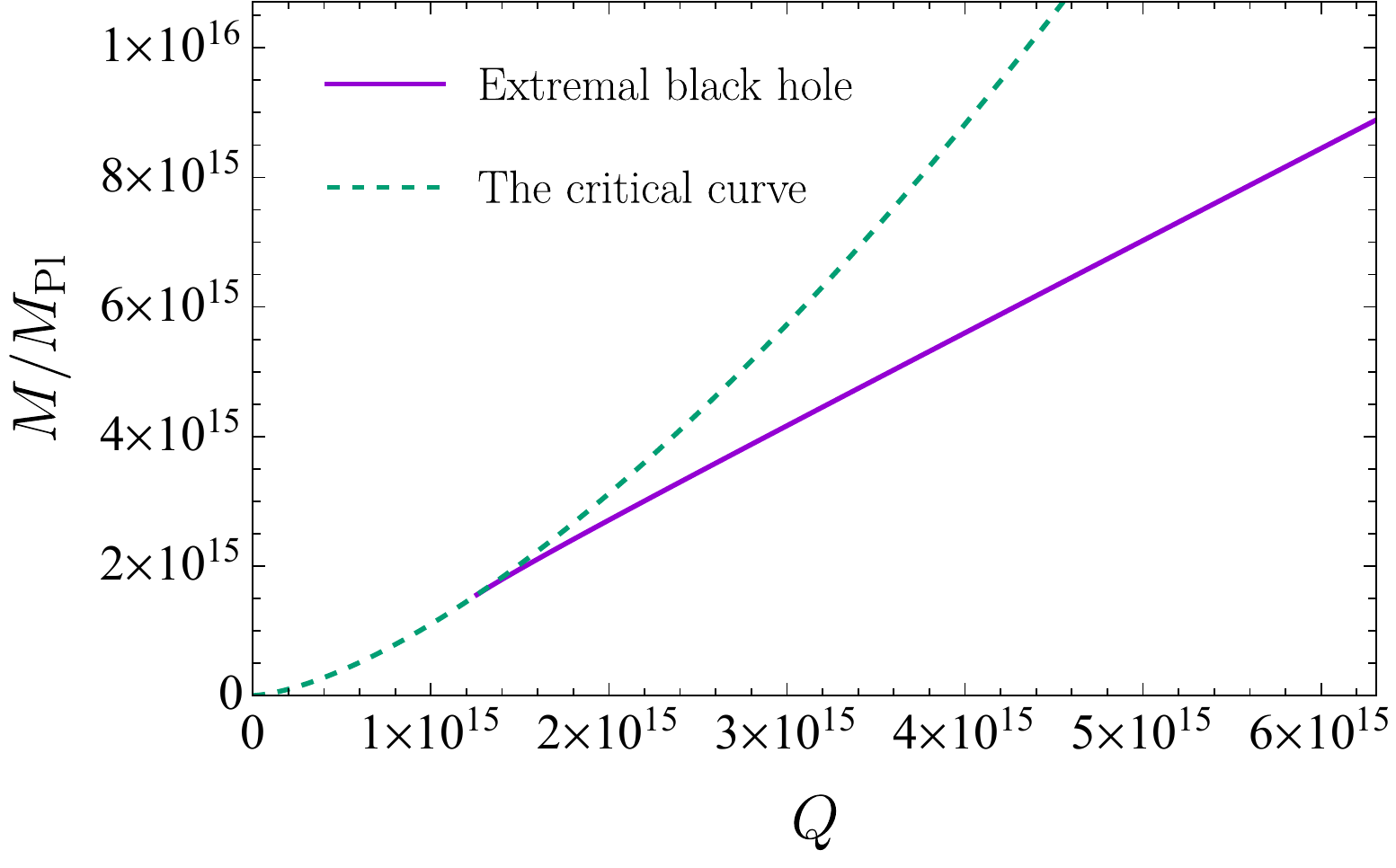}
     \caption{Phase diagram of DBI black holes in flat spacetime for $\Lambda_\DBI=10^{-7}\MPl$}
 \label{fig:DBI_phase_flat}
\end{figure}

As we mentioned, Eq.~\eqref{rH_DBI_flat} shows that the horizon degeneracy does not occur for $Q\leq (2G\Lambda_\DBI^2)^{-1}$, which is in sharp contrast to black holes in the Einstein-Maxwell theory. To elaborate on this feature, it is convenient to take a closer look at the shape of the function $f(r)$ defining the horizon. In particular, it turns out that the sign of $f(+0)$ is crucial. More explicitly, $f(r)$ behaves in the limit $r\to+0$ as
\begin{align}
\nonumber
	f(r)&\simeq  - \frac{2 G M}{r} + \frac{8 \pi G}{r} \int_\infty^0 dr' r'^2 \mc{L}\left(\tfrac{n^2}{8r'^4},0\right)
 \\*
 &=\frac{2G}{r}\left(M_{\rm crit.}(Q)-M\right)\,,
\end{align}
where we introduced the critical mass for a given charge $Q$ by
\begin{align}
M_{\rm crit.}(Q)&:=4\pi \int_\infty^0 dr' r'^2 \mc{L}\left(\tfrac{n^2}{8r'^4},0\right)
=
4\pi \int_\infty^0 dr' r'^2
\Lambda_\DBI^4 \left(
		1 - \sqrt{ 1 + \frac{Q^2}{16\pi^2\Lambda_\DBI^4r'^4}}\,
	\right)
\nonumber
\\*
\label{critical_mass}
&=\frac{3 Q^{3/2} \Lambda_\DBI}{64\pi}\left(\Gamma \left(-\frac{3}{4}\right)\right)^2\,.
\end{align}
In Fig.~\ref{fig: critical_flat}, we illustrate typical shapes of $f(r)$ below and above the critical mass $M_{\rm crit.}$. The upper four figures are for $Q> (2G\Lambda_\DBI^2)^{-1}$: (a) When the mass is below the mass $M_{\rm ext.}(Q)$ of the extremal black hole, there are no solutions for $f(r)=0$. (b) The extremal black hole has the degenerate horizon. (c) For $M_{\rm ext.}(Q) <M<M_{\rm crit.}(Q)$, there are two horizons. (d) If we increase the mass further, the sign of $f(0+)$ flips at the critical mass and as a consequence there exists only one positive real solution for $f(r)=0$. On the other hand, the lower two figures are for $Q< (2G\Lambda_\DBI^2)^{-1}$: (e) There is no positive real solution for $f(r)=0$ below the critical mass. (f) There is one above the critical mass. In particular, there is no regime with two horizons. In the small $Q$ region and above the critical curve, the horizon structure of the DBI black holes is similar to the Schwarzschild one. To summarize this feature, it is useful to draw a phase diagram given in Fig.~\ref{fig:DBI_phase_flat}. There, we find that the extremal curve and the critical curve intersect at $Q= (2G\Lambda_\DBI^2)^{-1}$.

\section{Black holes in dS and AdS}
\label{sec:4}

In this section we extend the flat space analysis of the previous section to black holes with a nonzero cosmological constant. We discuss asymptotically de Sitter (dS) spacetime in Sec.~\ref{sec:4.1} and then asymptotically anti-de Sitter (AdS) spacetime in Sec.~\ref{sec:4.2}.

\subsection{de Sitter black holes}
\label{sec:4.1}

Our task here is basically the same as the flat space case: We use the same algorithm to identify the condition~\eqref{eq:extremalcond_magnetic} for horizon degeneracy and then evaluate the corresponding black hole mass using the mass formula~\eqref{eq:Mmagnetic}. A new feature here is that there exists a cosmological horizon in addition to black hole horizons, so that our focus will be more on how the Nariai curve is modified in the nonlinear electrodynamics.

\subsubsection{Euler-Heisenberg black holes in de Sitter}

\begin{figure}[t]
 \centering
 \includegraphics[width=0.65\textwidth]{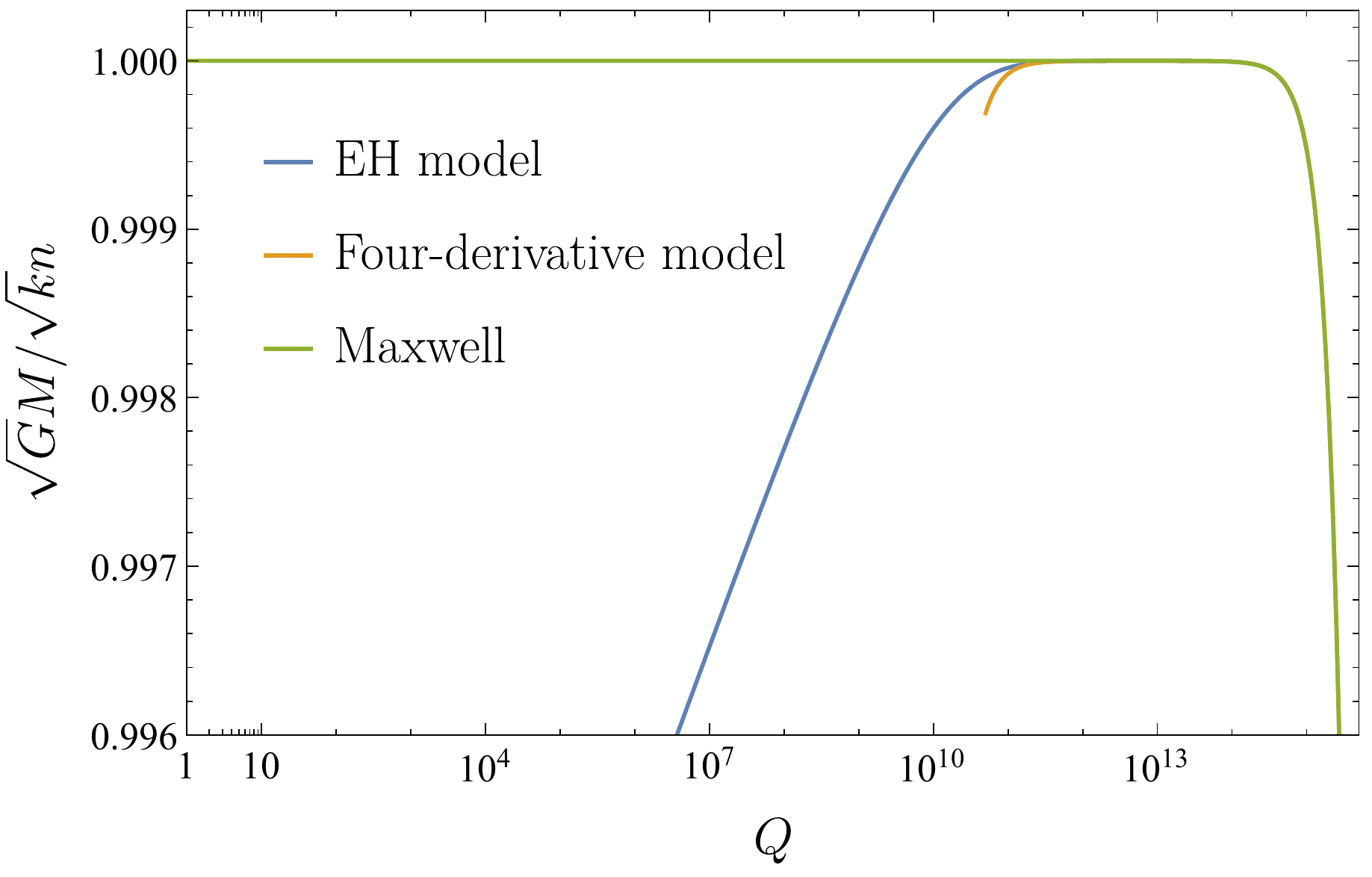}
 \quad
 \includegraphics[width=0.65\textwidth]{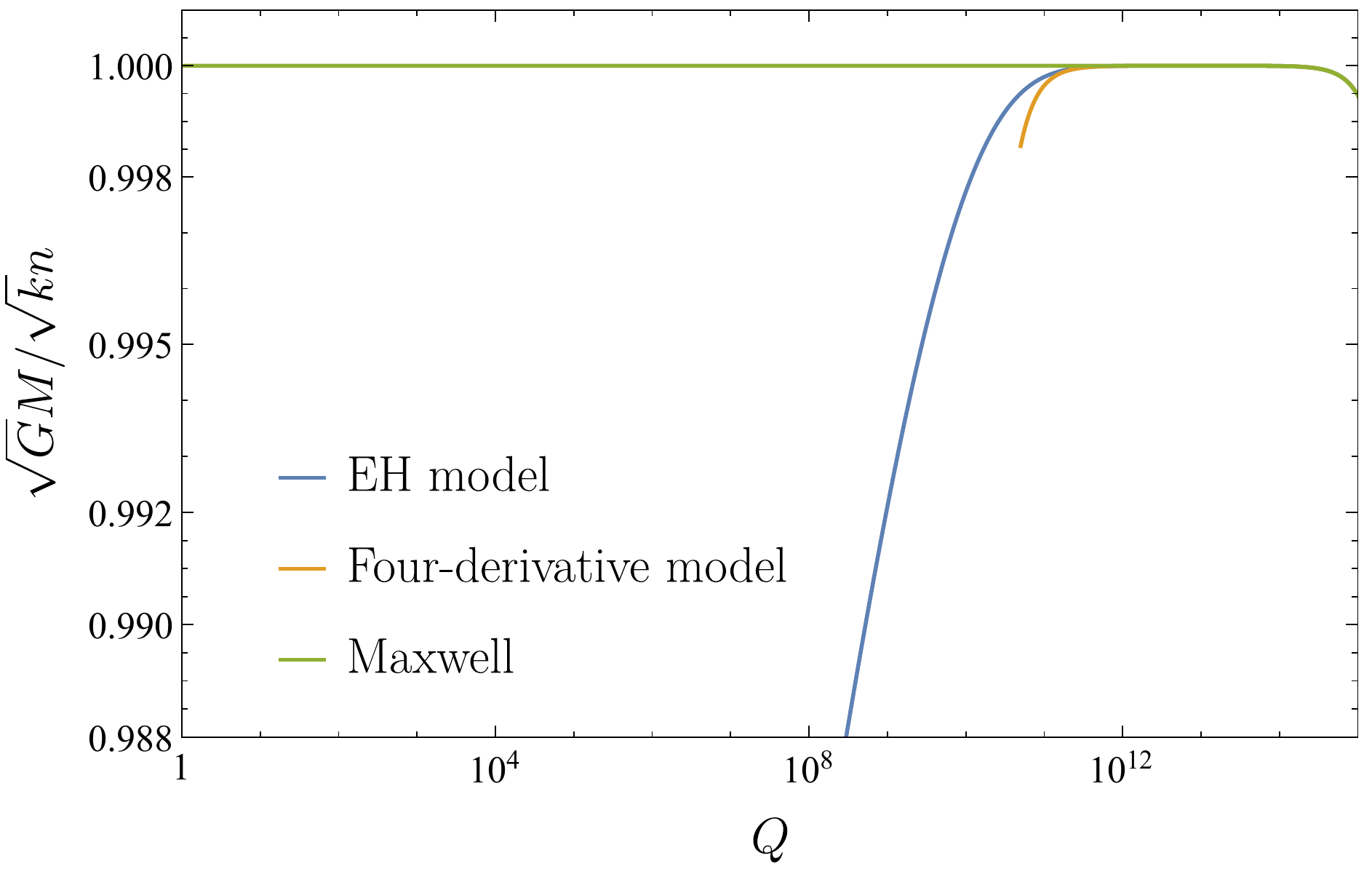}
 \caption{Extremal conditions of the EH model in dS spacetime for $\Lambda=(10^{-15}\MPl)^2$, $m=10^{-5}\MPl$ and $g_e=1$: The upper/lower figure is for scalar/fermion.
 }
 \label{fig:MtoQEHds}
\end{figure}

We begin by the EH model with a positive cosmological constant $\Lambda>0$. To avoid Schwinger effects, we consider magnetic black holes as before. Generically there exist two solutions for the condition~\eqref{eq:extremalcond_magnetic} for horizon degeneracy, where the smaller/larger horizon radius corresponds to the extremal/Nariai black hole. We solve the condition~\eqref{eq:extremalcond_magnetic} numerically and then evaluate the corresponding mass numerically using the formula~\eqref{eq:Mmagnetic}, drawing the extremal and Nariai curves.

\medskip
For illustration, we set the cosmological constant as $\Lambda=(10^{-15} \MPl)^2$. Fig.~\ref{fig:MtoQEHds} shows the mass-to-charge ratio of extremal black holes in the EH model for scalar/fermion loop (upper/lower) with the mass $m=10^{-5}\MPl$ (blue curve) in comparison with the Einstein-Maxwell theory (green curve) and the four-derivative model (orange curve). Recall that the charge-to-mass ratio of extremal black holes in the Einstein-Maxwell theory is not constant in dS because of the curvature effects. Therefore, what we expect is the monotonicity of the correction to the extremal condition, rather than the extremal curve itself (see also Fig.~\ref{fig:monotonicity_image}). Indeed we confirm the monotonicity of the correction in the EH model (and also in the four-derivative model). Besides, the correction in the EH model is milder than the four-derivative model, similarly to the flat space case.

\medskip
\begin{figure}[t]
 \centering
 \includegraphics[width=0.65\textwidth]{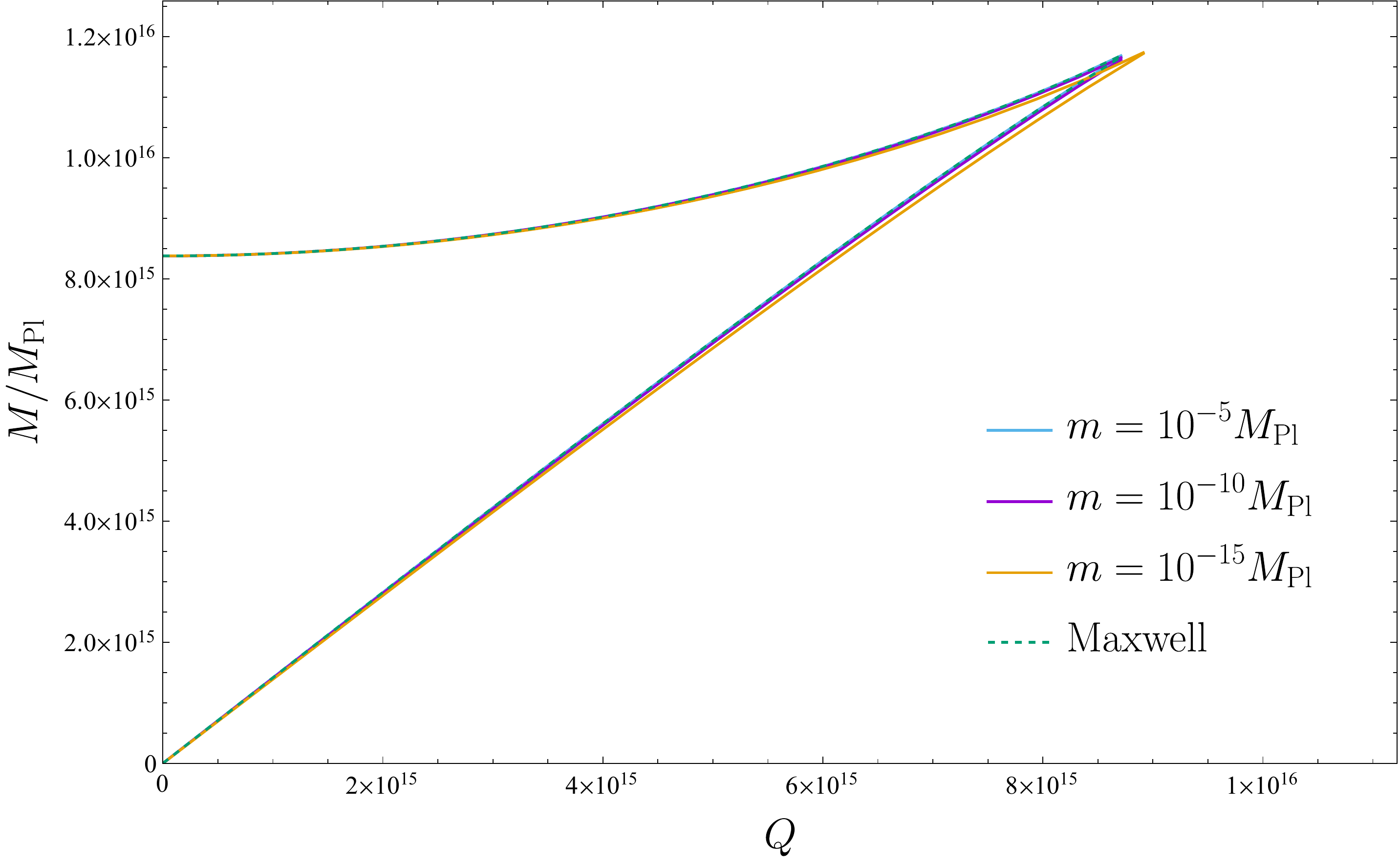}
 \quad
 \includegraphics[width=0.65\textwidth]{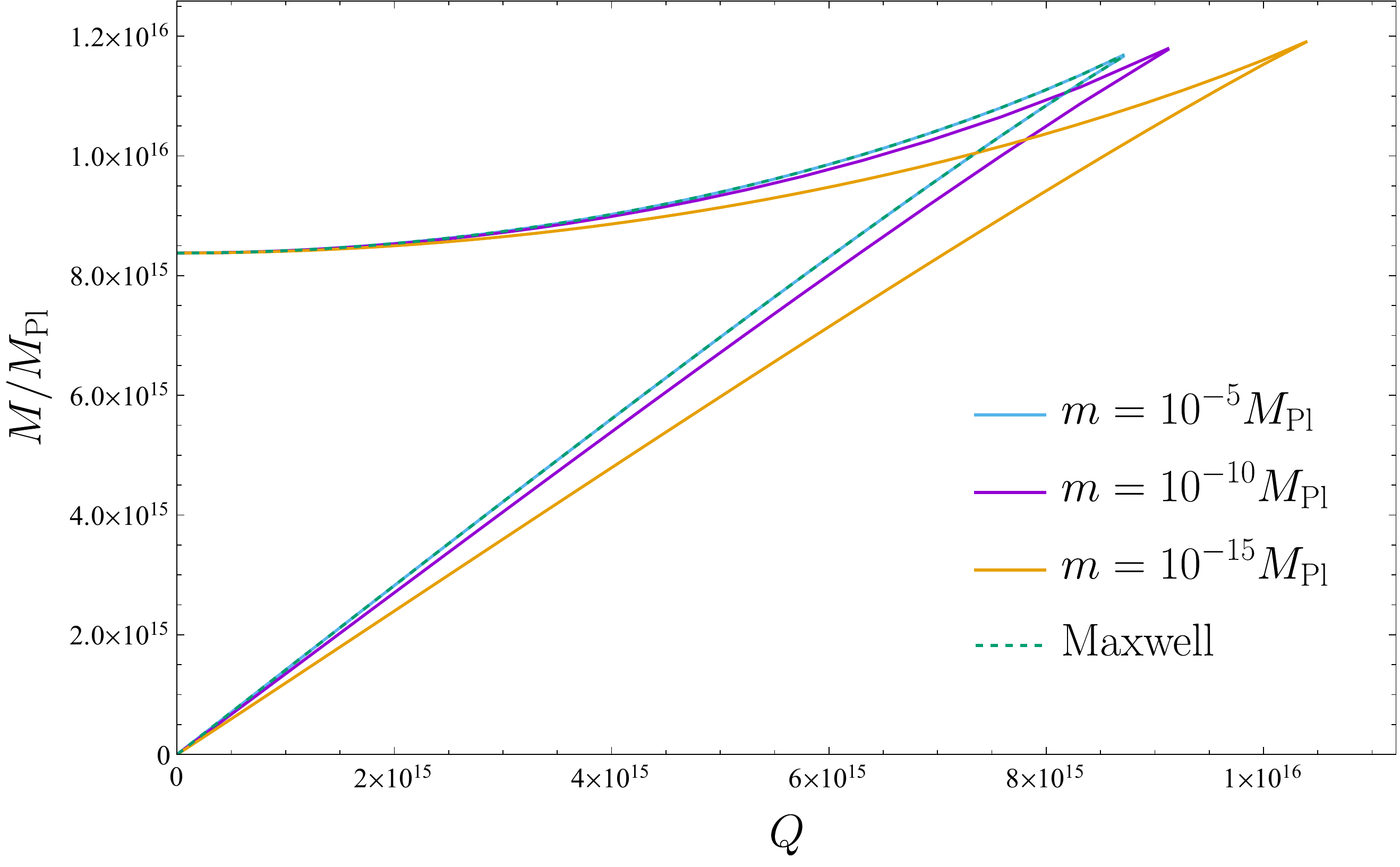}
 \caption{The extremal and the Nariai curves in dS spacetime with EH model for $\Lambda=(10^{-15}\MPl)^2$, $g_e=1$ and different masses ($m=10^{-5}\MPl$ (blue), $m=10^{-10}\MPl$ (purple) and $m=10^{-15}\MPl$ (orange)): The upper/lower figure is for scalar/fermion.}
 \label{fig:EH_shark}
\end{figure}

\medskip
In Fig.~\ref{fig:EH_shark}, we show how the ``shark fin" shape surrounded by the extremal curve and the Nariai curve is modified in the EH model for scalar/fermion loop (upper/lower) with the mass  $m = 10^{-5}\MPl$ (blue curve), $m = 10^{-10} \MPl$ (purple curve), and $m= 10^{-15} \MPl$ (orange curve),\footnotemark[5]\footnotetext[5]{
The EH model is applicable when the Compton length of the charged particle is smaller than the Hubble scale $m\gtrsim \Lambda^{1/2}$. Even though $m= 10^{-15} \MPl$ is marginal to this bound, we show the result for illustration.}
in comparison to the Einstein-Maxwell theory (dashed curve). We find that the Nariai curve is flattened by the nonlinear effects and this correction is larger for the lighter charged particle. We also find that this feature is more significant for the fermion loop than the scalar loop. 
Note that if we make the gauge coupling $g_e$ smaller, the shark fin shape approaches to the Einstein-Maxwell one similarly to the flat spacetime case.

\paragraph{validity of Euler-Heisenberg model in de Sitter.}

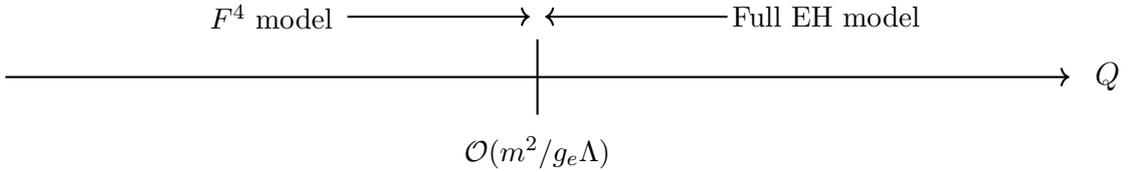
\begin{figure}[h]
	\centering
	\begin{tikzpicture}
		\draw[->,thick](-7,0)--(7,0);
		\draw[thick](0,-0.5)--(0,0.5);
		\draw(7.5,0)node{\large$Q$};
		\draw[->,thick](-2.5,0.8)--(-0.1,0.8);
		\draw[<-,thick](0.1,0.8)--(2.5,0.8);
		\draw(-3.5,0.8)node{$F^4$ model};
		\draw(3.8,0.8)node{Full EH model};
		\draw(0,-1)node{$\mathcal{O}(m^2/g_e \Lambda)$};
	\end{tikzpicture}
	\caption{
The full-order analysis of EH model beyond the four-derivative approximation is needed for $Q\gtrsim (m^2/g_e\Lambda)$.
	}
    \label{fig:validEHdS}
\end{figure}

Similar to the flat case, we go into detail on for which charge range the use of the EH Lagrangian~\eqref{eq:EH-Lagrangian-scalar} is justified and the full order analysis is needed in de Sitter case. First of all, in EH model charged particles are not dynamical. In de Sitter space time this condition means that charged particle are not excited by de Sitter background, which reads $m>\sqrt{\Lambda}$. For extremal black holes, the rest conditions are the same as the flat space case, but for Nariai black holes, the valid energy region is different. First, in EH model, the electromagnetic field $F_{\mu\nu}$ is assumed to be nearly constant at the Compton scale of the charged particle integrated out, which is given by $\frac{1}{m}\left|\frac{\del F}{\del r}\right|\ll |F|$. For Nariai magnetic black holes, this condition is satisfied when the compton length of charged particle integrated out is smaller than de Sitter radius. Also, we need full order analysis without four derivative truncation when the inequality $|\mc{F}|\gtrsim m^4$ is satisfied. For Nariai magnetic black holes, this condition reads $n^2 \Lambda^2\gtrsim m^{-4}$, which is equivalent to $Q\gtrsim \frac{m^2}{g_e \Lambda}$. See Fig.~\ref{fig:validEHdS} for a summary of the paragraph.

\paragraph{Implications for the FL bound.}

We close our EH analysis on de Sitter by discussing possible implications for the Festina Lente (FL) bound~\cite{Montero:2019ekk}, which was originally proposed based on thought experiments about decay of Nariai black holes: In the presence of electrically charged particles, electric Nariai black holes decay by emitting radiation due to Schwinger effects. If the discharge process is too fast compared to the energy loss, the black hole may decay into a naked singularity outside the shark fin. By postulating that this process is prohibited, Ref.~\cite{Montero:2019ekk} proposed a lower bound
\begin{align}
    m \gtrsim \sqrt{qgM_{\rm{Pl}} H}
\end{align}
on the mass of charged particles that has to be satisfied by all charged particles, where the Hubble constant $H$ is related to the positive cosmological constant $\Lambda$ as $H\sim \Lambda^{1/2}$ and $q$ is the integer charge of the particle. Now let us recall our results showing that light electrically charged particles may flatten the Nariai curve of magnetic black holes by the nonlinear effects of the EH model. We expect that a similar phenomenon will happen for electric black holes too. If it is indeed the case, we need to revisit the original FL argument about the Nariai black hole decay based on the black hole spectrum modified by backreaction from the light charged particles prohibited by the bound. We leave this issue for future work.

\medskip
A more direct relation of our analysis and the FL bound can be found along the line of the argument in Ref.~\cite{Huang:2006hc,Montero:2021otb} which fixed the $\mathcal{O}(1)$ coefficient of the bound as we summarize below: In four dimensions, the electric WGC for a unit charge requires existence of a charged particle satisfying the bound,
\begin{align}
\label{WGC_unit}
m\leq \sqrt{2}gM_{\rm Pl}\,.
\end{align}
Also, let us parameterize the $\mathcal{O}(1)$ coefficient of the FL bound as
\begin{align}
\label{FL_alpha}
m \geq \alpha\sqrt{qgM_{\rm{Pl}} H}
\end{align}
with an $\mathcal{O}(1)$ coefficient $\alpha$. Since the WGC particle has to satisfy the FL bound, combining Eq.~\eqref{WGC_unit} and Eq.~\eqref{FL_alpha} with $q=1$ gives
\begin{align}
\label{FL_like}
g\geq \frac{\alpha^2}{2}\frac{H}{M_{\rm Pl}}\,.
\end{align}
It is similar to the condition that magnetic black holes with a unit charge can exist in de Sitter spacetime, which is given by
\begin{align}
\label{eq:glower}
    g \geq \sqrt{\frac{3}{2}}\frac{H}{M_{\rm{Pl}}}
\end{align}
in the Einstein-Maxwell theory. Ref.~\cite{Huang:2006hc,Montero:2021otb} fixed the $\mathcal{O}(1)$ coefficient of the FL bound as $\alpha=6^{1/4}$ by postulating that the two conditions~\eqref{FL_like}--\eqref{eq:glower} match with each other. Now let us recall our analysis showing that the shark fin shape of magnetic black hole, especially the maximum charge of magnetic black holes, is modified in the presence of electrically charged particles. Therefore, the bound~\eqref{eq:glower} is modified in the presence of light charged particles prohibited by the FL bound and therefore the $\mathcal{O}(1)$ coefficient may be modified by such backreaction. It would be interesting to explore further along the line of this consideration to sharpen the FL bound.

\subsubsection{DBI black holes in de Sitter}

\begin{figure}[t]
 \centering
 \includegraphics[width=0.65\textwidth]{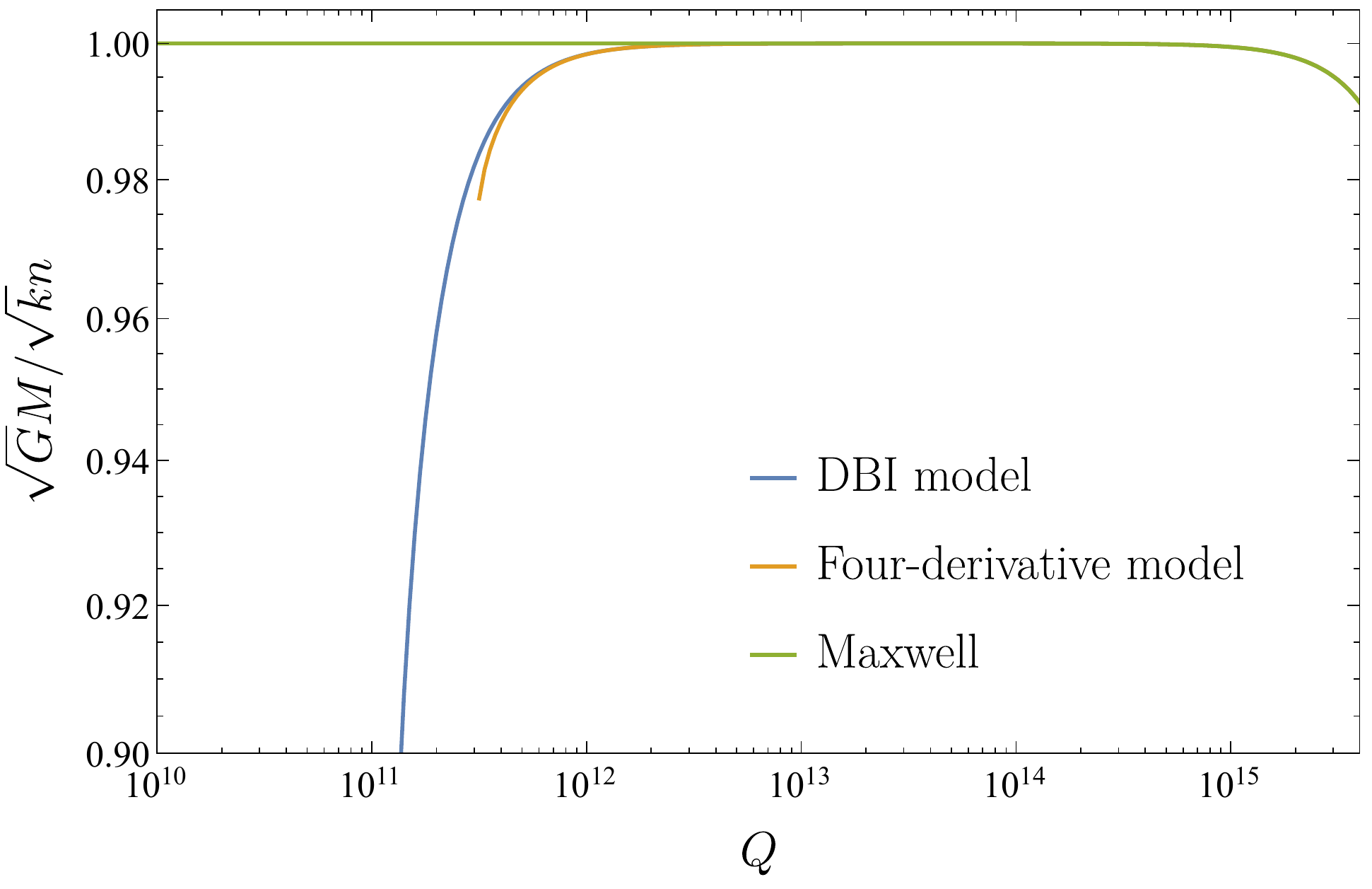}
 \caption{Extremal condition of DBI model in dS for $\Lambda=(10^{-15}\MPl)^2$, $\Lambda_\DBI=10^{-5}\MPl$.}
 \label{fig:DBIds}
\end{figure}

\begin{figure}[t]
 \centering
 \includegraphics[width=0.65\textwidth]{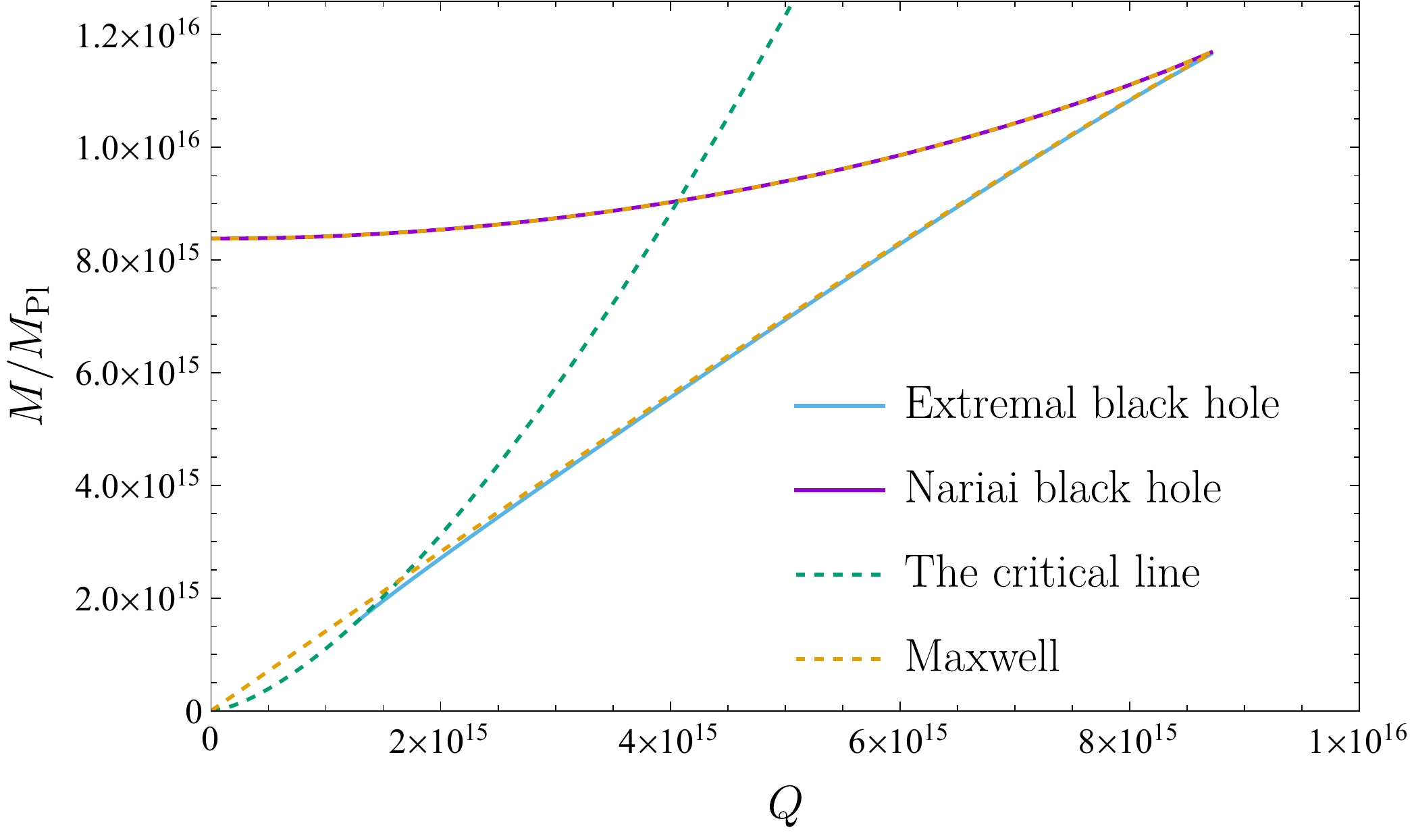}
 \quad
 \includegraphics[width=0.65\textwidth]{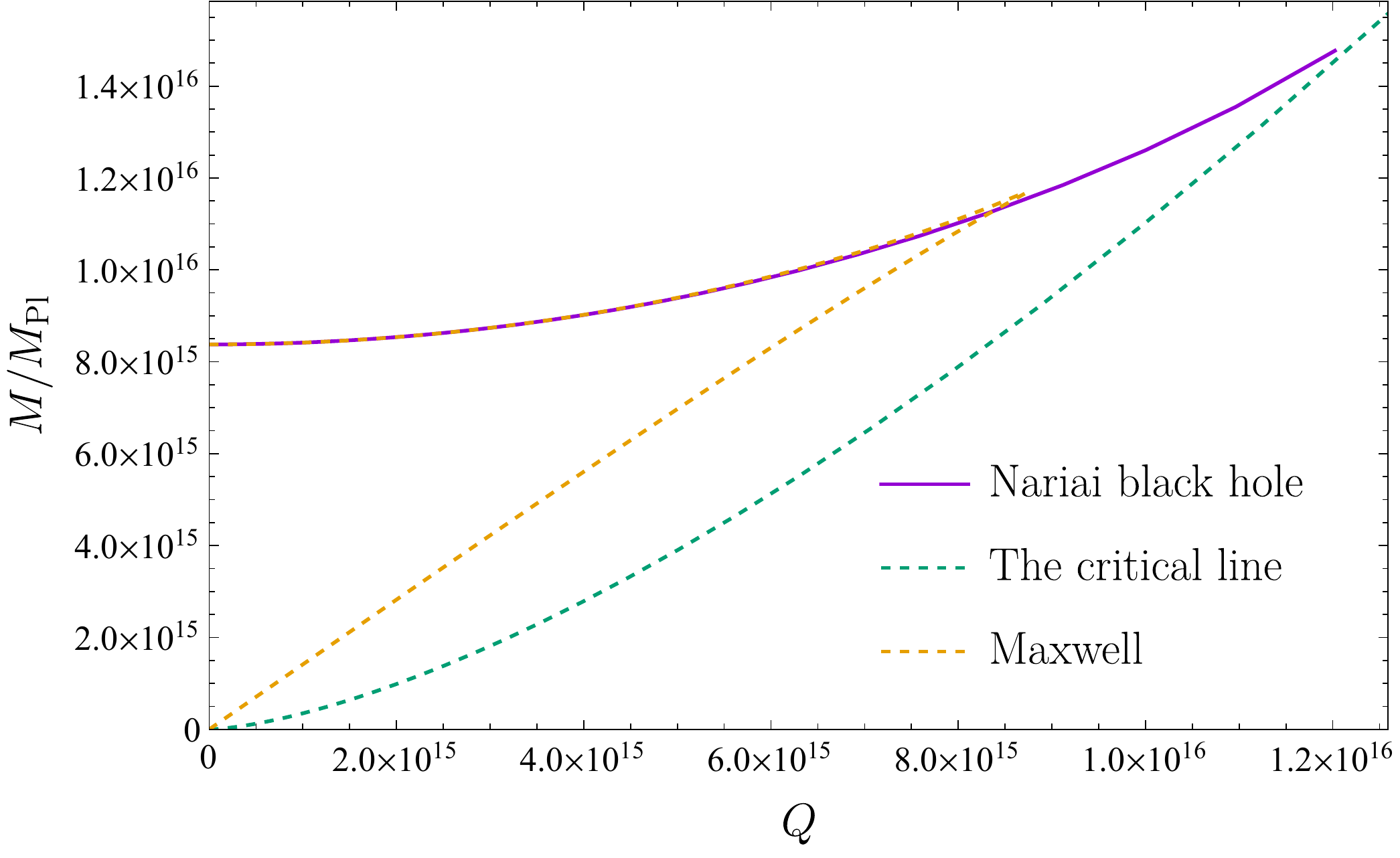}
 \caption{Spectrums of the magnetically charged black hole for DBI model in comparison to the Einstein-Maxwell model (orange dashed): The upper and lower panels are for $\Lambda_\DBI=10^{-7} \MPl$ and $\Lambda_\DBI=10^{-7.5} \MPl$. The blue/purple curve is for extremal/Nariai black holes and the critical curve is the green dashed one.
}
 \label{fig:DBIM_shark}
\end{figure}

Next we consider the DBI model with a positive cosmological constant $\Lambda>0$. Thanks to the electromagnetic duality of the DBI model, the analysis of  electric black holes is essentially the same as the magnetic case, so that we again focus on magnetic black holes. First, the condition~\eqref{extremal_DBI_general} for horizon degeneracy generically has two positive real solutions:
\begin{align}
r_H=\sqrt{\frac{8\pi G \Lambda_\DBI^4-\Lambda \pm 2 G \Lambda_\DBI^2\sqrt{16\pi^2\Lambda_\DBI^4+\Lambda^2Q^2-16\pi G\Lambda \Lambda_\DBI^4Q^2}}{\Lambda (16\pi G \Lambda_\DBI^4-\Lambda)}},
\end{align}
where the plus/minus sign corresponds to the Nariai/extremal condition. Note that the solution corresponding to the extremal condition can be used also in the AdS analysis, whereas that for the Nariai condition becomes complex when the cosmological constant is negative. Substituting it into Eq.~\eqref{eq:magneticDBIM} gives the mass-charge relation of the Nariai/extremal black holes. Also, similarly to the flat space case, there exists a critical mass~\eqref{critical_mass} beyond which the number of horizons changes because of the sign flip of $f(+0)$.

\medskip
For illustration, we again consider $\Lambda=(10^{-15}M_{\rm Pl})^2$. First, Fig.~\ref{fig:DBIds} shows the mass-to-charge ratio of extremal black holes in the DBI model with the mass $m=10^{-5}\MPl$. Similarly to the EH case, we confirm the monotonicity of the correction to the extremal condition and also find that the correction in the DBI model is milder than the four-derivative model. 

\medskip
Fig.~\ref{fig:DBIM_shark} shows how the shark fin structure is modified in the DBI model. For each region, the shape of $f(r)$ is similar to the flat space case, except that there is a cosmological horizon. The upper and lower panels are for the DBI scale  $\Lambda_\DBI = 10^{-7} \MPl$ and $\Lambda_\DBI = 10^{-7.5} \MPl$, respectively. For $\Lambda_\DBI = 10^{-7.5} \MPl$, extremal black holes do not exist. There is an event horizon in the intermediate region where $M_{\rm crit.}(Q) <M<M_{\rm Nariai.}(Q)$, so this is the allowed black hole region. The orange dashed curve shows the spectrum in the Einstein-Maxwell theory. The blue, purple and green dashed  curves are the extremal, Nariai, and critical curves, respectively. Each black hole on the critical curve has the minimum mass for the given charge.

\subsection{Anti-de Sitter black holes}
\label{sec:4.2}

\begin{figure}[h]

 \centering
 \includegraphics[width=0.6\textwidth]{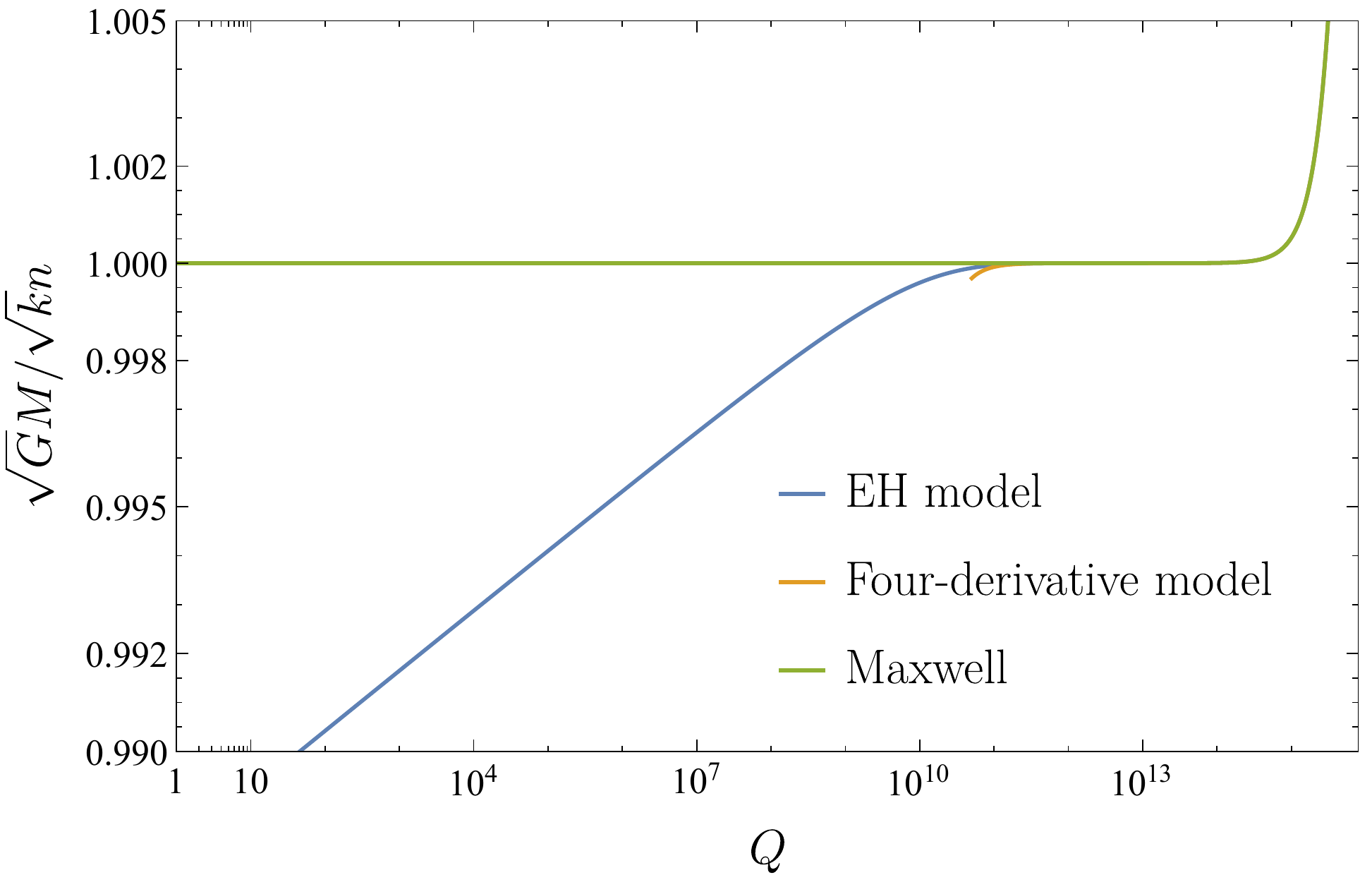}
 \quad
 \includegraphics[width=0.6\textwidth]{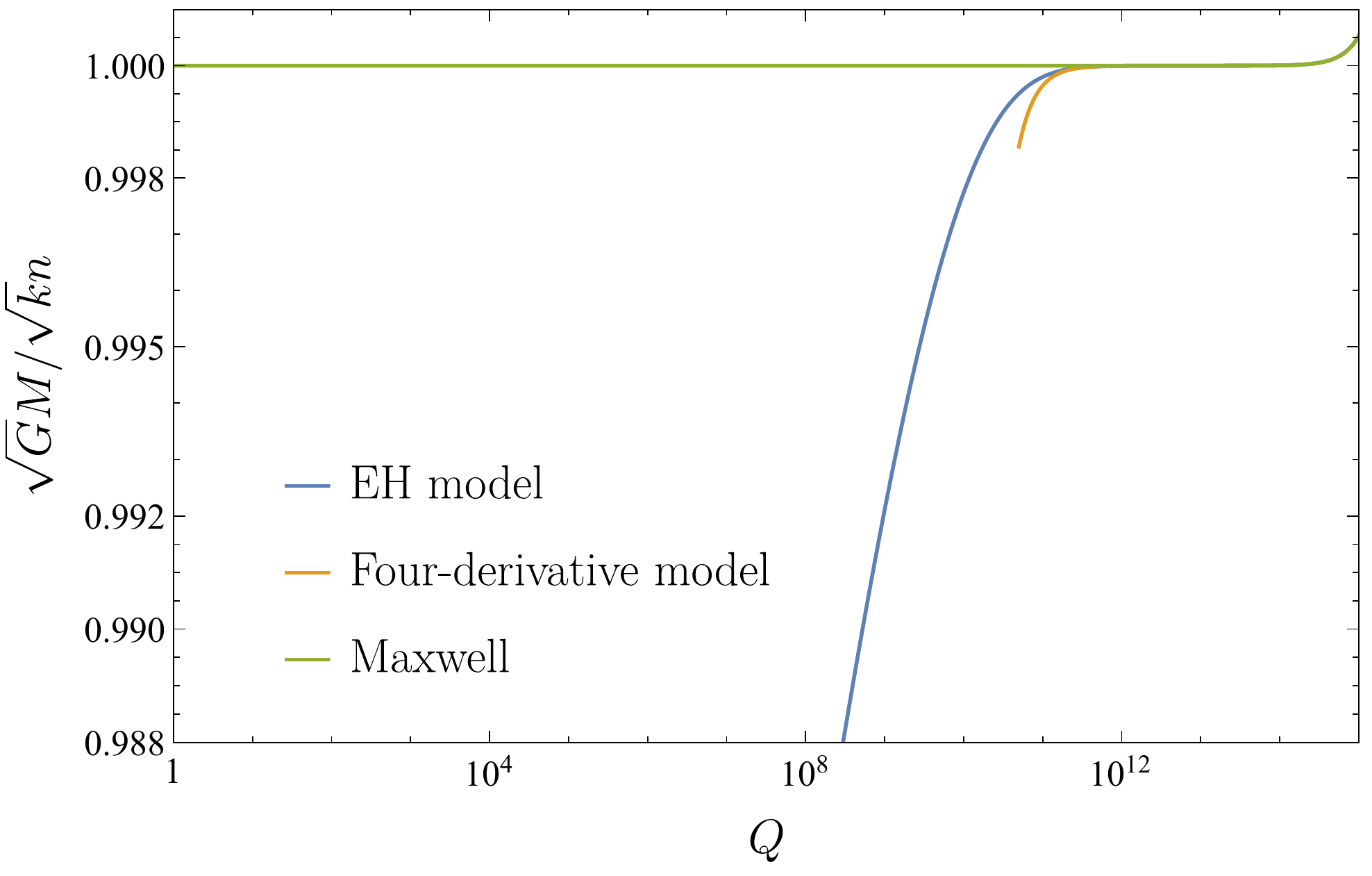}
 \caption{Extremal conditions of EH model in AdS for $\Lambda=-(10^{-15}\MPl)^2$ and $m=10^{-5}\MPl$: The upper/lower panel is for  scalar/fermion.
 }
 \label{fig:MtoQEHads}

\vspace{10mm}

 \centering
 \includegraphics[width=0.6\textwidth]{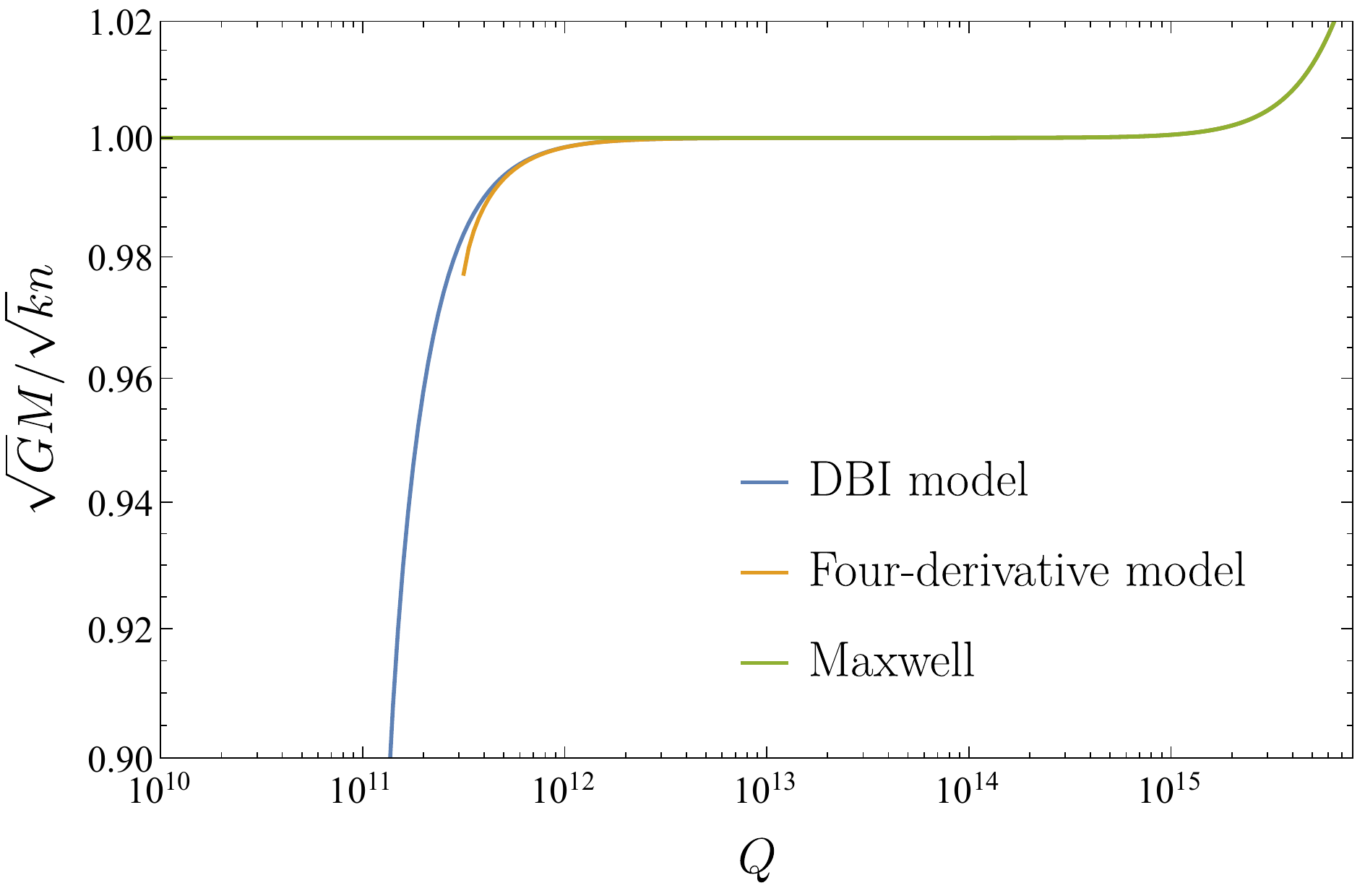}
 \caption{
 Extremal condition of DBI model in AdS for $\Lambda=-(10^{-15}\MPl)^2$ and $\Lambda_\DBI=10^{-5}\MPl$.
 }
 \label{fig:DBIads}
\end{figure}

Finally, we consider black holes in AdS. Since there is no cosmological horizon and therefore there is no Nariai black hole, the results for AdS are qualitatively similar to the flat space case. The algorithm to derive the extremal condition is the same as previous examples. Therefore we just provide final plots for the extremal condition. For illustration, we set the cosmological constant as $\Lambda = - (10^{-15} \MPl)^2$. Fig.~\ref{fig:MtoQEHads} and Fig.~\ref{fig:DBIads} show the mass-to-charge ratio of the extremal black holes in the EH model and the DBI model, respectively. There we confirm the monotonicity of the correction to the extremal condition.

\newpage
\section{Black hole analogue of gravitational positivity}
\label{sec:5}

In this section we point out an interesting similarity between our black hole analysis and positivity bounds on scattering amplitudes~\cite{Pham:1985cr,Adams:2006sv}, especially in gravity theories. In Sec.~\ref{sec:3.1} we evaluated the mass-to-charge ratio,
\begin{align}
\label{mu_ext}
\mu \coloneqq \frac{\sqrt{G}M}{\sqrt{k_m}n}\,,
\end{align}
of extremal magnetic black holes in the Euler-Heisenberg model with  $\Lambda=0$. While the Einstein-Maxwell theory provides a good approximation as long as the black hole charge is large enough, the nonlinearity becomes important once the charge becomes as small as the critical value $Q_*\sim g_e(\MPl/m)^2$ (see also Fig.~\ref{fig:validEH}). For sufficiently small $Q\ll Q_*$, the correction to the mass-to-charge ratio~\eqref{mu_ext} in the EH model scales logarithmically as
\begin{align}
\label{mu_log}
\Delta\mu_{\rm EH}\sim -g_e^2\ln\frac{Q_*}{Q}\,.
\end{align}
Here and in what follows we do not care about $\mathcal{O}(1)$ factors, even though we care the sign. Physically, the logarithmic behavior corresponds to the running of the gauge coupling induced by the charged particle. More quantitatively, the energy scale $E$ associated with the electromagnetic fields near the horizon reads
\begin{align}
\label{E_F}
    E \sim \mc{F}^{1/4} \sim \biggl( \frac{n^2}{r_H^4} \biggr)^{1/4} \sim \biggl( \frac{g_e^2 Q^2}{r_H^4} \biggr)^{1/4} \sim (g_e/Q)^{1/2}\MPl\,,
\end{align}
where we used $Q = g_m n$, $g_m \sim 1/ g_e$, and $r_H \sim Q/ \MPl$.
Hence we can think of $Q^{-1/2}$ as a measure of the energy scale in the Planck unit.

\medskip
While the EH model captures non-gravitational corrections to the Einstein-Maxwell theory from charged particles, there exist gravitational corrections as well. For example, four-derivative operators schematically of the form $F^2R$ are generated at one loop. Their contribution to the extremal condition is (see, e.g., Ref.~\cite{Hamada:2018dde})
\begin{align}
\Delta\mu_{\rm grav}\sim \frac{g_e^2\MPl^2}{m^2}Q^{-2}
\sim\frac{m^2}{M_{\rm Pl}^2}\left(\frac{Q_*}{Q}\right)^2\,,
\end{align}
where we emphasize that the gravitational correction is positive. In the spirit of the WGC, let us postulate that the mass-to-charge ratio of extremal black holes has to be smaller than unity. Then, we obtain the following bound\footnotemark[6]\footnotetext[6]{
Here we implicitly assumed that the charged particle satisfies the WGC bound, having QED in mind. Otherwise, the gravitational correction dominates over the non-gravitational one and then the total correction to the mass-to-charge ratio of extremal black holes become positive even in the large $Q$ region, where the four-derivative model is applicable.
}:
\begin{align}
\label{Q_bound}
\Delta\mu_{\rm EH}+\Delta\mu_{\rm grav}<0
\quad
\longleftrightarrow
\quad
\frac{Q}{Q_*}\gtrsim\frac{m}{g_e\MPl}\,.
\end{align}
Interestingly, the bound can be rephrased in terms of the energy scale~\eqref{E_F} as
\begin{align}
\label{E_bound}
E\lesssim \sqrt{g_em\MPl}\,,
\end{align}
which is reminiscent of the cutoff energy scale suggested by gravitational positivity bounds in QED~\cite{Alberte:2020bdz}.\footnotemark[7]\footnotetext[7]{
To be precise, positivity bounds in the presence of gravity hold only approximately, at least in the present technology. See~\cite{Hamada:2018dde,Bellazzini:2019xts,Alberte:2020jsk,Tokuda:2020mlf,Herrero-Valea:2020wxz,Caron-Huot:2021rmr,Alberte:2021dnj,Bellazzini:2021oaj,Caron-Huot:2022ugt,Chiang:2022jep,Herrero-Valea:2022lfd,deRham:2022gfe,Noumi:2022wwf,Hamada:2023cyt} for recent discussion. The cutoff scale $E\lesssim\sqrt{g_em\MPl}$ follows under the assumption that the allowed negativity does not dominate over the negative gravitational contribution.
}

\paragraph{Caveat.}

While the above observation is interesting and suggestive, a caveat is needed: When the bounds~\eqref{Q_bound}--\eqref{E_bound} are saturated, the black hole radius is comparable to the Compton length of the charged particle, $r_H\sim Q/\MPl\simeq g_e/m$, so that we cannot justify the use of the EH model. However, we expect that the logarithmic behavior~\eqref{mu_log} still holds even in this regime because it is related to the running of the gauge coupling. It would be desirable to reformulate our analysis in terms of running couplings from the Wilsonian EFT perspective, which we leave for future work.

\paragraph{Interpretation.}

Given this caveat, we interpret that requirement of the WGC type bound $\mu\leq 1$ for extremal black holes with arbitrary charge $Q$ provides the black hole analogue of improved positivity bounds~\cite{Bellazzini:2016xrt,deRham:2017avq,deRham:2017imi}:
To explain this, it is convenient to compare our EH analysis with the four-derivative analysis in the literature. As we explained in Sec.~\ref{sec:3.1}, the four-derivative analysis is valid only for sufficiently large charge $Q\gg Q_*$ or in other words only in the low-energy limit. On the other hand, our EH analysis is applicable even for smaller charge $Q\lesssim Q_*$, so that we can test the WGC type inequality $\Delta\mu=\Delta \mu_{\rm EH}+\Delta\mu_{\rm grav}\leq0$ for a wider range of $Q$. If we employ the WGC type bound $\Delta\mu\leq0$ as a criterion for consistent gravity theories, one may ask up to which value of $Q$ the bound is satisfied and how to modify the theory such that $\Delta\mu\leq0$ is satisfied for all $Q$. Since $Q$ is associated with energy, this is equivalent to identifying the cutoff scale and asking how to UV complete the theory. This is the same philosophy as the improved positivity bounds, which provide an energy-scale-dependent bound useful for identifying the cutoff scale. Indeed, our EH analysis implies the same cutoff scale as gravitational positivity in QED.

\section{Conclusion}
\label{sec:6}

In this paper, we studied the extremal condition of charged black holes in nonlinear electrodynamics beyond the four-derivative corrections. More specifically, we considered the Euler-Heisenberg model and the DBI model in asymptotically flat spacetime, de Sitter spacetime, and anti-de Sitter spacetime. In all cases, we confirmed the monotonicity of the correction to the mass-to-charge ratio of extremal black holes, which supports the black hole version of the Weak Gravity Conjecture. Our analysis took into account all orders in the derivative expansion, so that its applicability is not limited to the large black hole limit or in other words the low-energy limit. Indeed, we found that the corrections in the Euler-Heisenberg model and the DBI model are milder than the four-derivative model, which offers a concept of the UV completion in the black hole context.

\medskip
Our analysis for asymptotically de Sitter black holes is relevant to the Festina Lente bound too. We used the Euler-Heisenberg model to demonstrate that the Nariai curve for magnetic black holes is flattened by light (electrically) charged particles. This is relevant to the argument in Ref.~\cite{Montero:2021otb} that fixed the $\mathcal{O}(1)$ coefficient of the bound. Moreover, if a similar flattening by light charged particles happens for electric black holes, we need to revisit original discussion motivating the bound. It would be interesting to study electric black holes in the Euler-Heisenberg model, appropriately taking into account Schwinger effects captured by the imaginary part of the effective Lagrangian.

\medskip
Besides, we found an interesting similarity between our black hole analysis and positivity bounds on scattering amplitudes. In the spirit of the black hole WGC, we postulated that the mass-to-charge ratio of extremal magnetic black holes is smaller than unity $\mu\leq 1$ for arbitrary charge $Q$ and then the Euler-Heisenberg analysis implied a cutoff energy scale $\sim \sqrt{g_emM_{\rm Pl}}$ similar to the one implied by gravitational positivity bounds in QED~\cite{Alberte:2020bdz}. This observation would be useful when sharpening positivity bounds in the presence of gravity. It would be interesting to collect more evidences for the correspondence in more realistic models along the line of Refs.~\cite{Alberte:2020bdz,Aoki:2021ckh,Noumi:2022zht,Aoki:2023khq}. Such an interplay between the black hole thermodynamics and the S-matrix bootstrap would broaden our global view of the bootstrap in gravity theories.

\section*{Acknowledgments}
We would like to thank Kimihiro Nomura for friendly, encouraging and detailed discussion about black holes in nonlinear electrodynamics. We also thank Yu-tin Huang, Keisuke Izumi, Gary Shiu, Pablo Soler and Daisuke Yoshida for useful discussion.
The work of Y.A. is supported by JSPS Overseas Research Fellowships.
T.N. is supported in part by JSPS KAKENHI Grant No. 20H01902 and No. 22H01220, and MEXT KAKENHI Grant No. 21H05184 and No. 23H04007.

\newpage
\appendix

\section{Details of numerical analysis}
\label{app:AppendixA}

\begin{figure}[t]
    \centering
      \includegraphics[width=0.48\textwidth]{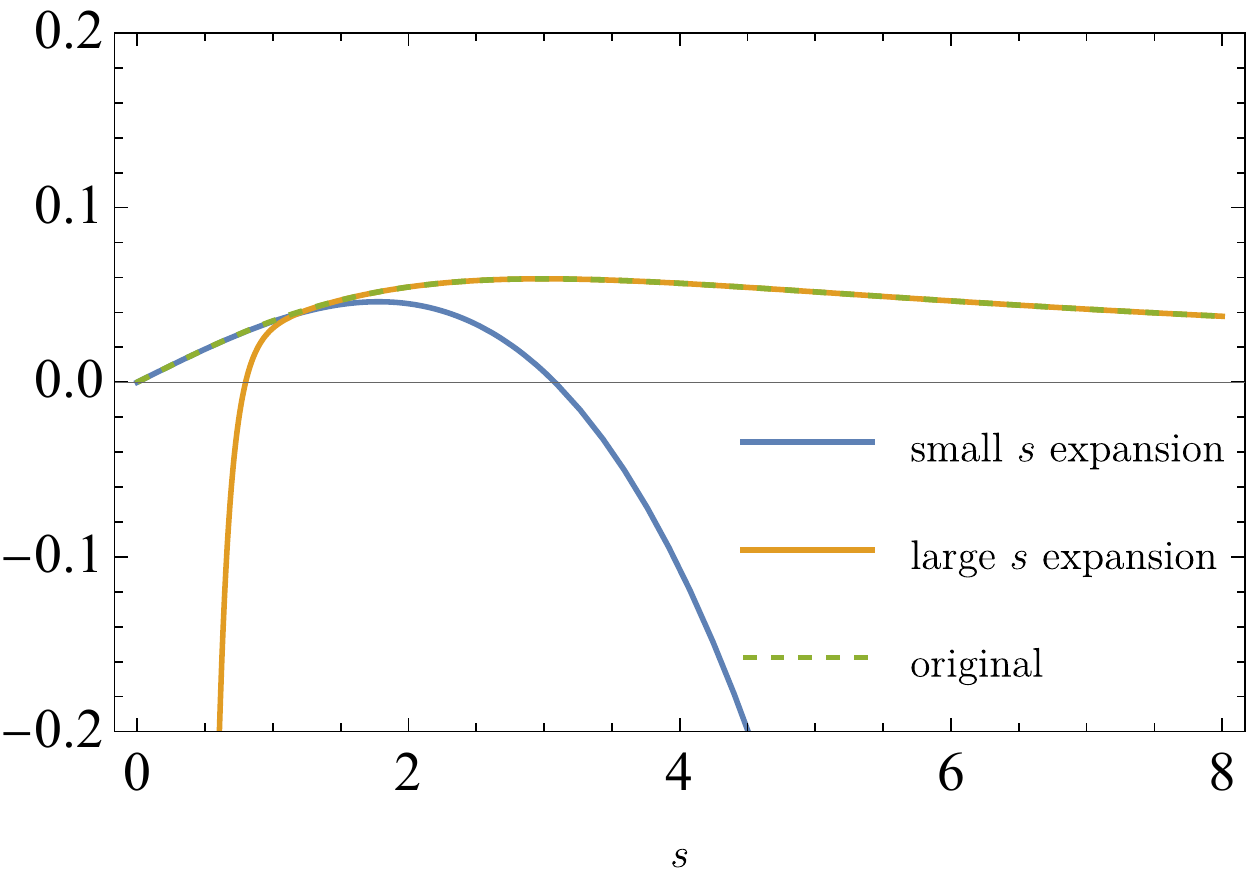}
      \quad 
      \includegraphics[width=0.48\textwidth]{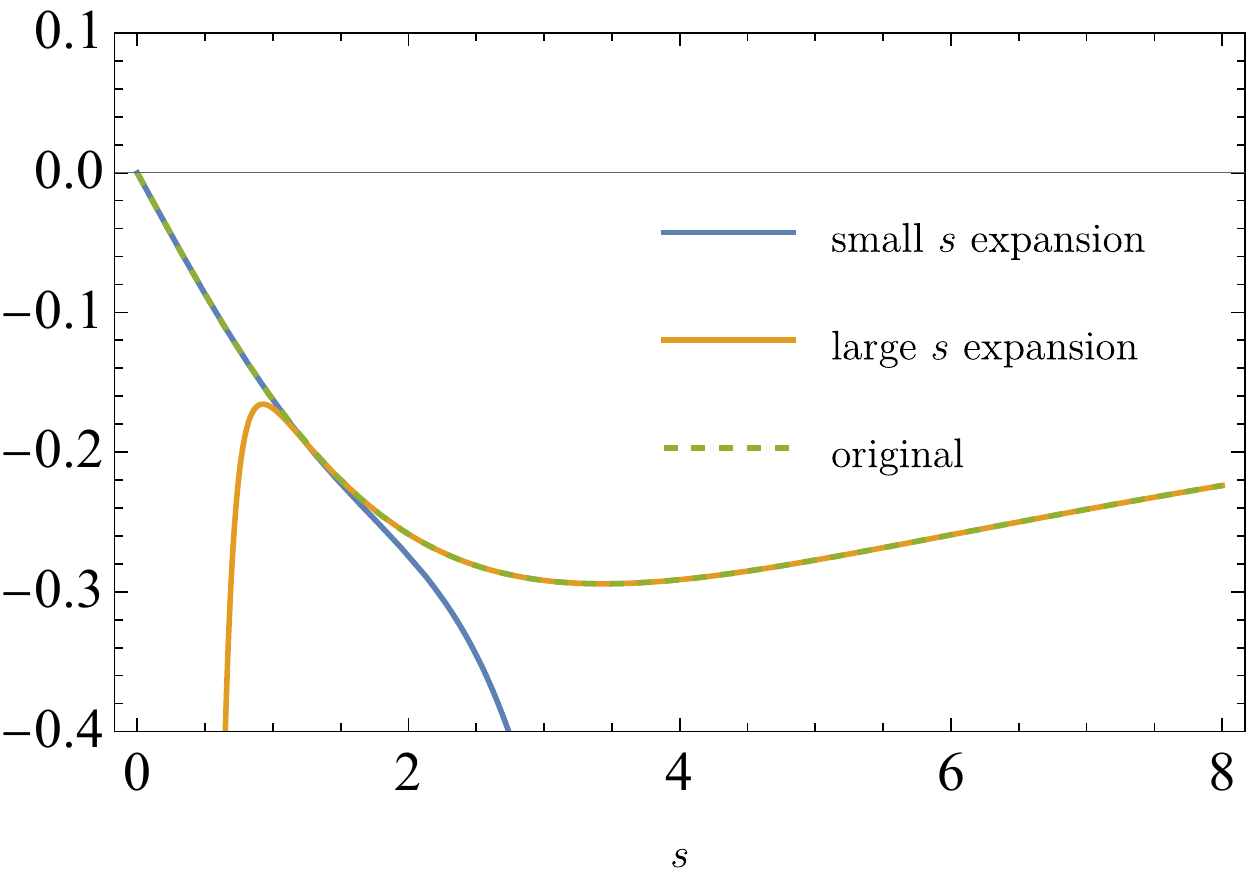}
    \caption{(Left) scalar
    (Right) fermion}
    \label{fig:integrand}
\end{figure}

We provide technical details on the approximation and numerical calculation used in the Euler-Heisenberg analysis. First, in order to derive an analytic approximation of the EH Lagrangian~\eqref{eq:EH-Lagrangian}, we divide the integration range of the second term into $0\leq s \leq 1$ and $1\leq s \leq \infty$, and perform Taylor expansion of the integrand  in each range. In the range $0\leq s\leq1$, we expand the integrand in $s$ around $s=0$ up to the fourth/sixth order for scalar/fermion loop. On the other hand, in the range  $1 \leq s \leq \infty$, we expand in $e^{-s}=0$ around $e^{-s}=0$ ($s=\infty$) up to the fifth/third order for scalar/fermion loop. In Fig.~\ref{fig:integrand} the original integrand (green dashed), the expansion around $s=0$ (blue), and that around $s=\infty$ (orange) are compared for the parameter choice $\frac{\sqrt{\mc{F}}}{m^2}=10^{5}$. There we find a good agreement between the original integrand and our analytic approximation in each range.

\medskip
Under this approximation, we find an analytic form of the Euler-Heisenberg Lagrangian $\mc{L}(\frac{n^2}{8r^4},0)=\mc{L}(\frac{Q^2}{8g_m^2r^4},0)$ as follows: For scalar loop,  
\begin{small}
\begin{align}
    \label{eq:scEHcalculation}
    \mathcal{L}&=-\frac{Q^2}{32 \pi^2 r^4}+\frac{m^4 Q^2}{256 \pi^2 r^4 g_m^2 m^4}\left[\sum_{l=1}^3 4e^{-(2l-1)-\frac{2 g_m m^2 r^2}{Q}}-\frac{31 Q^4}{20160 g_m^4 m^8 r^8}+\frac{7 Q^2}{720 g_m^2 m^4 r^4}\right.\nonumber\\*
    &\quad
    -\frac{e^{-\frac{2 g_m m^2 r^2}{Q}}\left(Q-2 g_m m^2 r^2\right)}{Q}+\frac{e^{-\frac{2 g_m m^2 r^2}{Q}}Q\left(-1052 g_m^3 m^6 r^6-402 g_m^2 m^4 Q r^4+186 g_m m^2 Q^2 r^2+93 Q^3\right)}{60480 g_m^4 m^8 r^8}\nonumber\\*
    &\quad
+\frac{4 g_m^2 m^4 r^4 \Ei\left(-\frac{2 g_m m^2 r^2}{Q}\right)}{Q^2}+\frac{1}{3}\Gamma\left(0, \frac{2 g_m m^2 r^2}{Q}\right)\nonumber\\*
    &\quad
\left.-\sum_{l=1}^3\frac{4 \left(2 g_m m^2 r^2+(2l-1)Q\right) \Gamma \left(0,\frac{2 g_m m^2 r^2}{Q}+2l-1\right)}{Q}\right]\,.
\end{align}
\end{small}

\noindent
For fermion loop,
\begin{small}
\begin{align}
\label{eq:feEHcalculation}
    \mathcal{L}&=-\frac{Q^2}{32 \pi^2 r^4}+\frac{m^4 Q^2}{256 \pi^2 r^4 g_m^2 m^4}\left[4e^{-\frac{2g_m m^2 r^2}{Q}}+\sum_{l=1}^{3}16\left(e^{-2l-\frac{2 g_m m^2 r^2}{Q}}\right)-\sum_{l=1}^{3} 32 n\Gamma\left(0,2l+\frac{2 g_m m^2 r^2}{Q}\right)\right.\nonumber\\*
    &\quad
-\frac{32 g_m m^2 r^2 \sum_{l=1}^{3} \Gamma \left(0,2l+\frac{2 g_m m^2 r^2}{Q}\right)}{Q}-\frac{Q^6}{315 g_m^6 m^{12} r^{12}}+\frac{e^{-\frac{2 g_m m^2 r^2}{Q}}Q^6}{315 g_m^6 m^{12} r^{12}}\nonumber\\*
    &\quad
+\frac{2 e^{-\frac{2 g_m m^2 r^2}{Q}}Q^5}{315 g_m^5 m^{10} r^{10}}+\frac{2 Q^4}{315 g_m^4 m^{8} r^{8}}-\frac{8 e^{-\frac{2 g_m m^2 r^2}{Q}}Q^3}{945 g_m^3 m^6 r^6}-\frac{2 Q^2}{45 g_m^2 m^4 r^4}+\frac{32 e^{-\frac{2 g_m m^2 r^2}{Q}}Q^2}{945 g_m^2 m^4 r^4}+\frac{128 e^{-\frac{2 g_m m^2 r^2}{Q}}Q}{1575 g_m m^2 r^2}\nonumber\\*
    &\quad
\left.+\frac{8 e^{-\frac{2 g_m m^2 r^2}{Q}}g_m m^2 rr^2}{Q}+\frac{16 g_m^2 m^4 r^4 \Ei\left(-\frac{2 g_m m^2 r^2}{Q}\right)}{Q^2}-\frac{8\left(Q+6 g_m m^2 r^2 \Gamma\left(0,\frac{2 g_m m^2 r^2}{Q}\right)\right)}{3 Q}\right]\,,
\end{align}
\end{small}

\noindent
where
\begin{align}
    \Gamma(a,z)=\int^\infty_z dt~t^{a-1}e^{-t},
\end{align}
and
\begin{align}
    \Ei(x)=\int^\infty_{-z} dt \frac{e^{-t}}{t}~.
\end{align}

\medskip
When we compute the black hole mass using the formula~\eqref{eq:Mmagnetic}, we need to perform integration of the Lagrangian. In order to improve the accuracy of the numerical calculation, we use the second order expansion of $r$ in the small $r$ region. Under this approximation, $\mathcal{L}$ is described as follows: For scalar one-loop,
\begin{small}
\begin{align}
    \mc{L}&=-\frac{Q^2}{32 \pi^2 r^4}+\frac{m^4 Q^2}{256 \pi^2 r^4 g_m^2 m^4}\left[\frac{4\left(1+e^2+e^4\right)}{e^5}-\frac{29683}{30240}-\frac{\gamma}{3}-4\sum_{l=1}^3(2l-1)\Gamma(0,2l-1) \right.\nonumber\\
    &\quad
\left.-\frac{1}{3}\log\left(\frac{2 g_m m^2}{Q}\right)-\frac{2}{3}\log(r)+\frac{g_m m^2 r^2}{Q}\left(\frac{9749}{2100}-8\sum_{l=1}^3\Gamma(0,2l-1)\right)\right]\,.
\end{align}
\end{small}

\noindent
For fermion one-loop,
\begin{small}
\begin{align}
    \mc{L}&=-\frac{Q^2}{32 \pi^2 r^4}+\frac{m^4 Q^2}{256 \pi^2 r^4 g_m^2 m^4}\left[8\left(\frac{2\left(1+e^2+e^4\right)}{e^6}+\frac{991}{2025}+\frac{\gamma}{3}-4\sum_{l=1}^3 (2l-1)\Gamma(0,2l)+\frac{1}{3}\log\left(\frac{2 g_m m^2}{Q}\right)\right.\right.\nonumber\\
    &\quad
\left.\left.+\frac{2}{3}\log(r)\right)+\frac{16 g_m m^2r^2}{Q}\left(-\frac{10793}{33075 g_m m^2}+\gamma-2\sum_{l=1}^3\Gamma(0,2l)+\log\left(\frac{2 g_m m^2}{Q}\right)+2\log(r)\right)\right]\,,
\end{align}
\end{small}

where 
\begin{align}
    \gamma\simeq 0.577216~.
\end{align}

\section{Anti-symmetric symbols and tensors}
\label{app:def-of-epsilon}

In this appendix, we summarize our convention of the anti-symmetric tensors and symbols.
Although we give the explicit form in four-dimensional case, it is straight forward to extend it to the general $D$-dimension.

\medskip
First, let us introduce the vierbein (vielbein) by 
\begin{align}
    g_{\mu\nu} = \eta_{ab} e^a{}_\mu e^b{}_\nu\,,
\end{align}
where $\mu, \nu, \ldots$ are the indices of the curved geometry and $a, b, \ldots$ denote those of the local Lorentz frame.
We use the mostly plus convention for the Minkowski metric $\eta_{ab} = \diag(-1, +1, \ldots ,+1)$.
The volume factor is 
\begin{align}
    \sqrt{-g} = \sqrt{-\det g_{\mu\nu}} = \det e^a{}_\mu \eqqcolon e\,.
\end{align}
On the local Lorentz frame, we introduce the anti-symmetric tensor as
\begin{align}
    \epsilon^{0123} = \alpha\,
    \qquad 
    \epsilon_{0123} = - \alpha
    \label{eq:epsilon-normalization}
\end{align}
up to a normalization factor $\alpha$.
Its standard choice is $\alpha = \pm 1$.\footnotemark[8]\footnotetext[8]{
    In this paper, we choose $\alpha = -1$.
}
In addition, let us introduce $\epsilon^{\mu\nu\rho\sigma}$ and $\epsilon_{\mu\nu\rho\sigma}$ by the action of the vierbein on these anti-symmetric tensor as 
\begin{align}
    \epsilon^{\mu\nu\rho\sigma} \coloneqq \epsilon^{abcd} e_a{}^\mu e_b{}^\nu e_c{}^\rho e_d{}^\sigma\,,
    \qquad 
    \epsilon_{\mu\nu\rho\sigma} \coloneqq \epsilon_{abcd} e^a{}_\mu e^b{}_\nu e^c{}_\rho e^d{}_\sigma\,.
\end{align}
They are related by raising and lowering the indices by the metric tensor.
The anti-symmetric symbols are defined by using them as 
\begin{align}
    \varepsilon^{\mu\nu\rho\sigma} \coloneqq e \epsilon^{\mu\nu\rho\sigma}\,,
    \qquad 
    \varepsilon_{\mu\nu\rho\sigma} \coloneqq e^{-1} \epsilon_{\mu\nu\rho\sigma}\,,
\end{align}
and they take the values of $0$ and $\pm 1$.
We note that these symbols are not related by raising and lowering the indices and the values are independent of the metric.
They satisfy the following relations and normalizations in $D$-dimension:
\begin{align}
    &\epsilon^{a_1 \cdots a_D} \epsilon_{a_1 \cdots a_D} = \varepsilon^{\mu_1 \cdots \mu_D} \varepsilon_{\mu_1 \cdots \mu_D} = - D! \alpha^2\,,
    \\
    & \epsilon^{a_1 \cdots a_p c_1 \cdots c_{D-p}} \epsilon_{b_1 \cdots b_p c_1 \cdots c_{D-p}} = -p! (D-p)! \alpha^2 \delta^{a_1 \cdots a_p}_{b_1 \cdots b_p}\,,
    \\
    &\varepsilon^{\mu_1 \cdots \mu_p \rho_1 \cdots \rho_{D-p}} \varepsilon_{\nu_1 \cdots \nu_p \rho_1 \cdots \rho_{D-p}} = - p! (D-p)! \alpha^2 \delta^{\mu_1 \cdots \mu_p}_{\nu_1 \cdots \nu_p}\,,
\end{align}
where we introduced 
\begin{align}
    \delta^{a_1 \cdots a_p}_{b_1 \cdots b_p} \coloneqq \frac{1}{p!} \sum_{\sigma \in S_p} \sgn(\sigma) \delta^{a_1}_{\sigma(b_1)} \cdots \delta^{a_p}_{\sigma(b_p)}\,.
\end{align}
$S_p$ denotes the $p$-dimensional symmetric group and $\sigma$ is its element.
$\sgn(\sigma)$ is the signature of the element $\sigma$.
The overall minus signature comes from the normalization \eqref{eq:epsilon-normalization}.
In addition, it is worth commenting on the following relation:
\begin{align}
    d x^\mu \wedge d x^\nu \wedge d x^\rho \wedge d x^\sigma = \frac{1}{\alpha} \varepsilon^{\mu\nu\rho\sigma} d^4x\,.
\end{align}

\medskip
In this paper, the dual field strength is defined by 
\begin{align}
    \tilde{F}^{\mu\nu} \coloneqq \frac{1}{2} \epsilon^{\mu\nu\rho\sigma} F_{\rho\sigma}\,.
\end{align}
We note that $\mc{G} = \frac{1}{4} F_{\mu\nu} \tilde{F}^{\mu\nu}$ depends on the metric only through the volume factor $\sqrt{-g}$ and the variation by the metric is 
\begin{align}
    \delta_g \mc{G} = \frac{1}{2} \mc{G} g_{\mu\nu} \delta g^{\mu\nu}.
\end{align}

\paragraph{Hodge dual.}

The Hodge dual of the $p$-form field $\omega_p$ is given by using the above anti-symmetric symbol as 
\begin{align}
    \ast_D \omega_p = \frac{e}{p! (D-p)!} \omega^{\nu_1 \cdots \nu_p} \varepsilon_{\nu_1 \cdots \nu_p \mu_1 \cdots \mu_{D-p}} d x^{\mu_1} \wedge \cdots \wedge d x^{\mu_{D-p}}\,,
\end{align}
where $\ast_D$ denotes the $D$ dimensional Hodge star.
This satisfies the following equations:
\begin{align}
    \ast_D \ast_D \omega_p = \alpha^2 (-1)^{p(D-p)+1} \omega_p\,,
    \quad
    \ast_D 1 = \frac{e}{D!} \varepsilon_{\mu_1 \cdots \mu_D} d x^{\mu_1} \wedge \cdots \wedge d x^{\mu_D} = - \alpha \sqrt{-g} d^Dx\,.
\end{align}
Using the Hodge star, the kinetic term of the $p$-form field is written as 
\begin{align}
    - \int d^Dx \sqrt{-g} \frac{1}{2} \frac{1}{(p+1)!} F_{\mu_1 \cdots \mu_{p+1}} F^{\mu_1 \cdots \mu_{p+1}} = \frac{1}{2\alpha} \int F_{p+1} \wedge \ast_D F_{p+1}\,,
\end{align}
where $F_{p+1} = d \omega_p$.

\section{Schwinger effect}
\label{Schwinger_effect}

In this paper we focus on magnetic black holes in order to ignore the Schwinger effect. Here we explain why the Schwinger effect cannot be neglected for electric black holes in the regime of our interests by comparing the charge loss rate with the black hole radius.

\medskip
Near the black hole horizon, the pair production rate per unit volume is 
\begin{align}
\Gamma=2\Im \mc{L} = \frac{g_e^2 Q^2}{64\pi^3 r_H^4} \sum^\infty_{n=1}\frac{1}{n\pi^2}\exp \left(-\frac{4 n\pi^2 m^2 r_H^2}{g_e Q}\right)\,.
\end{align}
Since the charge loss rate per unit volume is $-g_e\Gamma$, the charge loss rate of the black hole is estimated as
\begin{align}
\dot{Q}\sim -g_e\Gamma \,r_H^3\,.
\end{align}
The corresponding decay rate of the black hole charge reads
\begin{align}
\Bigg|\frac{\dot{Q}}{Q}\Bigg|
\sim
 \frac{g_e^3 Q^2}{r_H} \sum^\infty_{n=1}\frac{1}{n\pi^2}\exp \left(-\frac{4 n\pi^2 m^2 r_H^2}{g_e Q}\right)\,.
\end{align}
The Schwinger effect is negligible if it is sufficiently small compared to the curvature scale $r_H^{-1}$. Below, we examine this condition for extremal and Nariai black holes, respectively.


\paragraph{Extremal black hole.}
The curvature of the extremal black hole with the charge $Q$ is $\frac{1}{r_H}\sim \frac{1}{\sqrt{Gk}n}\sim\frac{\MPl}{Q}$, so that the decay rate of the black hole charge is
\begin{align}
    \frac{\dot{Q}}{Q}\sim -g_e^3 \MPl\sum^\infty_{n=1}\frac{1}{(n\pi)^2}\exp \left(-\frac{n Q m^2}{8 g_e \MPl^2}\right)~.
\end{align}
Therefore, we can ignore the Schwinger effect when 
\begin{align}
     g_e^3\MPl \sum^\infty_{n=1}\frac{1}{(n\pi)^2}\exp \left(-\frac{n Q m^2}{8 g_e \MPl^2}\right) \ll \frac{\MPl}{Q}\,.
\end{align}
Since the black hole charge is large $Q\gg 1$, this condition is satisfied only when the exponential suppression is sufficiently large. This means $Q\gg g_e \frac{\MPl^2}{m^2}$, which is nothing but the regime where the four-derivative model is applicable (see Fig.~\ref{fig:validEH}). In other words, the Schwinger effect is no more negligible for electric black holes with the charge $Q\lesssim g_e \frac{\MPl^2}{m^2}$, for which the nonlinear effect of the EH model (which is our main focus in the present paper) comes in.

\paragraph{Nariai black hole.}
The curvature of the Nariai black hole is $\frac{1}{r_H}\sim \sqrt{\Lambda}$, so that the decay rate of the black hole charge  is 
\begin{align}
    \frac{\dot{Q}}{Q}\sim-g_e^3 \MPl\sum^\infty_{n=1}\frac{1}{(n\pi)^2}\exp \left(-\frac{4\pi^2 n m^2}{g_e Q \Lambda }\right)~.
\end{align}
We can ignore the Schwinger effect when
\begin{align}
    g_e^3\MPl \sum^\infty_{n=1}\frac{1}{(n\pi)^2}\exp \left(-\frac{n Q m^2}{8 g_e \MPl^2}\right) \ll \sqrt{\Lambda}\,.
\end{align}
This condition is satisfied only when the exponential suppression is sufficiently large since $\sqrt{\Lambda}\ll \MPl$ (which we assume to treat gravity semiclasically). This means $Q \ll \frac{m^2}{g_e \Lambda}$, which is the regime where the four-derivative model is applicable. Thus for electric black holes with the charge $Q \gtrsim \frac{m^2}{g_e \Lambda}$, we cannot ignore the Schwinger effect.
In addition, the condition $Q \ll \frac{m^2}{g_e \Lambda}$ for the maximum Nariai black hole reads $m\gg (g_e \MPl \sqrt{\Lambda})^{1/2}$, which satisfies the FL bound. In other words, the Schwinger effect must be taken into account appropriately to study the Nariai curve in the presence of light charged particles that violate the FL bound: a careful study of the Schwinger effect will be required to sharpen the FL bound.

\newpage
\newcommand{\arxivfont}{\rmfamily}
\bibliographystyle{yautphys}
\bibliography{ref-EH}

\end{document}